\documentclass[12pt,preprint]{aastex}

\lefthead{Chou et al.}
\righthead{Metallicity of the Sagittarius Stream}

\begin{document}
\title{A 2MASS All-Sky View of the Sagittarius Dwarf Galaxy:
V. Variation of the Metallicity
Distribution Function Along the Sagittarius Stream}

\author{Mei-Yin Chou\altaffilmark{1},
Steven R. Majewski\altaffilmark{1}, Katia Cunha\altaffilmark{2,3},
Verne V. Smith\altaffilmark{2}, Richard J.
Patterson\altaffilmark{1}, David Mart{\'i}nez-Delgado
\altaffilmark{4}, David R. Law \altaffilmark{5}, Jeffrey D. Crane
\altaffilmark{1,6}, Ricardo R. Mu\~{n}oz\altaffilmark{1},
Ram\'{o}n Garcia L\'{o}pez \altaffilmark{7}, Doug
Geisler\altaffilmark{8}, and Michael F. Skrutskie\altaffilmark{1}}

\altaffiltext{1}{Dept. of Astronomy, University of Virginia,
Charlottesville, VA 22904-4325 (mc6ss, srm4n, rjp0i, rrm8f,
mfs4n@virginia.edu)}

\altaffiltext{2}{National Optical Astronomy Observatories, PO Box
26732, Tucson, AZ 85726 (cunha, vsmith@noao.edu)}

\altaffiltext{3}{On leave from Observatorio Nacional, Rio de
Janeiro, Brazil}

\altaffiltext{4}{Instituto de Astrophysica de Andalucia (CSIC),
Granada,Spain (ddelgado@iaa.es)}

\altaffiltext{5}{Department of Astronomy, California Institute of
Technology, MS 105-24, Pasadena, CA 91125
(drlaw@astro.caltech.edu)}

\altaffiltext{6}{Carnegie Observatories, 813 Santa Barbara Street,
Pasadena, CA 91101 (crane@ociw.edu)}

\altaffiltext{7}{Instituto de Astrofisica de Canarias, 38200 La
Laguna, Spain (rgl@ll.iac.es)}

\altaffiltext{8}{Departamento de Fisica, Universidad de
Concepci\'{o}n, Casilla 160-C, Concepci\'{o}n, Chile
(doug@kukita.cfm.udec.cl)}

\begin{abstract}

We present reliable measurements of the metallicity distribution
function (MDF) at different points along the tidal stream of the
Sagittarius (Sgr) dwarf spheroidal (dSph) galaxy, based on high
resolution, echelle spectroscopy of candidate M giant members of
the Sgr system. The Sgr MDF is found to evolve significantly from
a median [Fe/H] $\sim -0.4$ in the core to $\sim-1.1$ dex over a
Sgr leading arm length representing $\sim$2.5-3.0 Gyr of dynamical
(i.e. tidal stripping) age.  This is direct evidence that there
can be significant chemical differences between current dSph
satellites and the bulk of the stars they have contributed to the
halo. Our results suggest that Sgr experienced a significant
change in binding energy over the past several Gyr, which has
substantially decreased its tidal boundary across a radial range
over which there must have been a significant metallicity gradient
in the progenitor galaxy. By accounting for MDF variation along
the debris arms, we approximate the MDF Sgr would have had several
Gyr ago. We also analyze the MDF of a moving group of M giants we
previously discovered towards the North Galactic Cap having
opposite radial velocities to the infalling Sgr leading arm stars
there and propose that most of these represent Sgr {\it trailing
arm} stars overlapping the Sgr leading arm in this part of the
sky. If so, these trailing arm stars further demonstrate the
strong MDF evolution within the Sgr stream.

\end{abstract}

\keywords{galaxies: evolution -- Local Group -- galaxies: interactions -- Galaxy: halo}

\section{Abundances in Dwarf Galaxies and the Halo}

The idea that the stellar halo of the Milky Way (MW) formed {\it
predominantly} through the infall of smaller star systems ---
presumably dwarf galaxies --- has a long history (Searle \& Zinn
1978), strong observational evidence (e.g., Majewski 1993,
Majewski, Munn \& Hawley 1996), and currently a strong theoretical
backing by way of hierarchical, $\Lambda$CDM models (e.g., Bullock
\& Johnston 2005; Robertson et al. 2005; Abadi et al.\ 2006; Font
et al.\ 2006). But a longstanding puzzle in this picture is why,
if they are the seeds of halo formation, do MW satellite galaxies
have different stellar populations (e.g., Unavane, Wyse \& Gilmore
1996) and chemical abundance patterns (e.g., Fulbright 2002;
Shetrone et al.\ 2003; Tolstoy et al.\ 2003; Venn et al.\ 2004;
Geisler et al.\ 2005) than typical MW halo stars?  One explanation
(Majewski et al.\ 2002; Font et al.\ 2006) is that prolonged tidal
disruption will naturally lead to evolution in the types of stars
a particular satellite contributes to a halo. Indeed, it has
become clear that abundance patterns (e.g., [$\alpha$/Fe]) among
the most metal-poor stars in dSphs --- possibly the residue of a
formerly much larger metal-poor population that may have been
predominantly stripped from the satellites over their lifetime ---
do overlap those of halo stars of the same metallicity (Shetrone
et al.\ 2003; Geisler et al.\ 2005; Tolstoy 2005). But the true
connection of these ancient dSph stars with Galactic halo stars
remains speculative, or at least non-definitive.

The Sagittarius (Sgr) dSph provides a striking example of a
satellite galaxy being disrupted and slowly assimilated into the
MW halo field population.  It is the primary contributor of both
carbon stars and M giants to the upper ($|Z_{GC}| > 10$ kpc) halo
(Ibata et al.\ 2001; Majewski et al.\ 2003, hereafter Paper I) and
yields strong overdensity signatures of MSTO and RR Lyrae stars at
halo distances (Newberg et al.\ 2002; Vivas, Zinn \& Gallart
2005). Yet the current Metallicity Distribution Function (MDF) of
the Sgr core, with median [Fe/H] $\sim -0.4$ (Fig.\ 7), is quite
unlike that of the Galactic halo (median [Fe/H]$=-1.6$) and thus
the Sgr system would seem to present one of the most dramatic
examples of the apparent dSph/halo star abundance dichotomy.  In
this paper we explore the possible origins of this dichotomy by
making high resolution, spectroscopic observations of stars not
only known to have been contributed to the Milky Way halo from a
{\it specific} dSph satellite, but also {\it when}.  In the case
of the Sgr dSph we show that the origin of the abundance dichotomy
with the Galactic halo arises from preferential tidal stripping of
metal poor stars, which leads to divergent MDFs between lost and
retained Sgr stars, as well as a significant variation in the Sgr
MDF along its tidal tails from the core to debris lost from the
core several Gyr ago.

\section{Previous Abundance Studies of the Sgr System}

Initial photometric estimates indicated that Sgr is largely
dominated by a population of old to intermediate age stars
(Bellazzini et al.\ 1999; Layden \& Sarajedini 2000), but with an
MDF spanning from [Fe/H] $\sim -2.0$ to $\sim-0.5$ (see also
Cacciari, Bellazzini \& Colucci 2002). However, a more metal-rich
population with [Fe/H] $\geq -0.5$ was found with high resolution
spectra (Bonifacio et al.\ 2000, 2004; Smecker-Hane \& McWilliam
2002; Monaco et al.\ 2005) as well as in a recent, deep
color-magnitude diagram (CMD) from the Hubble Space Telescope ACS
centered on M54 (Siegel et al. 2007). These chemical abundance
studies thus present a Sgr MDF dominated by a metal-rich
population with median [Fe/H] $\sim -0.4$, but having a metal-weak
tail extending to [Fe/H] $\sim -2.0$ (Smecker-Hane \& McWilliam
2002; Zaggia et al.\ 2004; Monaco et al.\ 2005).  Monaco et al.\
(2003) and Cole et al.\ (2005) have found Sgr to have a similar
MDF to the LMC (which has a dominant population of median
[Fe/H]$=-0.4$) with a similar fraction of metal-poor stars, which
suggests that Sgr may have had a progenitor resembling the LMC
(Monaco et al.\ 2005).  In a recent reanalysis of the
age-metallicity relationship in Sgr, Bellazzini et al. (2006) find
that the dSph may have enriched to near-solar metallicity as early
as 6 Gyr ago, though a more recent analysis by Siegel et al.
(2007) suggests a somewhat slower evolution to this enrichment
level.

Thus far, abundance studies of the Sgr tails have been less
detailed. Dohm-Palmer et al.\ (2001) obtained spectra of some K
giants apparently in the northern leading arm (near its
apogalacticon) and inferred the stream there was about a half dex
more metal poor than the Sgr core; these authors suggested that
the Sgr dSph may originally have had a strong metallicity
gradient.  Alard (2001) noted differences in the Sgr giant branch
position in the $(J-K_s,K_s)_o$ CMD between the Sgr center and a
field $7\fdg5$ down its major axis implying a $-0.2$ dex
metallicity variation between these two points (see \S7). Paper I
also suggested the possibility of a metallicity variation along
the Sgr tidal arms because giant stars in the arms with different
$(J-K_s)_o$ colors seemed to yield different photometric parallax
distances for the stream when the color-magnitude relation of the
Sgr core was used for all colors; the differences could be
explained by varying mean RGB slopes along the stream (see Figure
14 and Footnote 14 of Paper I). Adding information derived from
isochrone-fitting to main sequence turnoff stars,
Mart{\'i}nez-Delgado et al. (2004) argued that there is a
substantial metallicity gradient along the Sgr stream. Vivas et
al.\ (2005) obtained a mean [Fe/H] $=-1.77$ from low/medium
resolution spectra of sixteen RR Lyrae stars in the Sgr leading
arm; but since only the oldest and hence metal-poor populations in
Sgr would produce RR Lyrae, this age-biased sample cannot be used
to infer information on the full extent of the stream MDF.  On the
other hand, Bellazzini et al. (2006) found significant differences
in the relative numbers of blue horizontal branch to red clump
stars between the Sgr core and a position about 75$^{\circ}$
forward along the Sgr leading arm, an imbalance that suggests a
significant metallicity variation along the Sgr stream.  Thus,
while compelling evidence has been gathering for metallicity
variations along the Sgr stream, no {\it direct} measurement of
this variation has been made by sampling with high resolution
spectroscopy the actual [Fe/H] distributions of constituent stars.

\section{Observations}
\subsection{Sample Selection}

We have begun a systematic survey of the chemical abundance patterns of
stars in the Sgr stream.
The goal of the present contribution is a first systematic exploration of the MDF along the
Sgr stream; future work will focus on chemical abundance {\it patterns} in Sgr stream stars.

The design of our study, and in particular the rationale for the
specific stars targeted for observation, has been driven by
several practical considerations.  First, because information on
potential {\it variations} in metallicity along the stream is
sought, multiple portions of the Sgr stream representing different
dynamical ages (i.e. the times when the debris was stripped) is
needed.  Second, because the Sgr core itself exhibits a
metallicity {\it spread}, insufficient information is gained by
only sampling a few stars at any particular part of the tail;
rather, exploration of {\it distributions} in metallicity is
needed.  This requires reasonable numbers of stars at each sampled
section of the stream.  With a limited amount of telescope time it
is easier to build large samples with brighter targets, but, even
focusing purely on the intrinsically brightest stars identified in
the stream --- the M giants explored, e.g., in Paper I and
Majewski et al. (2004, ``Paper II'' hereafter), this is still a
challenging project if spectra at echelle resolution are needed.
The difficulty of securing large samples of stars partly motivated
our strategy in this first study of Sgr debris stars to explore
the Sgr leading arm --- which passes quite near the solar
neighborhood (Paper I).  In contrast, the Sgr trailing arm, in its
most clearly discernible parts in the southern Galactic
Hemisphere, never gets closer than $\sim$15 kpc to the Sun.  By
observing the leading arm both just above and just below the
Galactic plane we access two different points along this tidal
stream with fairly local stars bright enough to take maximal
advantage of our particular instrument access (two echelle
spectrographs on 4-m class telescope in the Northern Hemisphere
and only about one night per year on an echelle spectrograph in
the Southern Hemisphere).

This strategy for exploring the leading arm, however, has some
drawbacks in that (1) the trailing arm is dynamically much better
understood than the leading arm (Helmi 2004; Law, Johnston \&
Majewski 2005, hereafter ``Paper IV''), (2) the sorting of stars
by dynamical age is much cleaner in the trailing arm than the
leading arm (Paper IV; see also \S 7), (3) major sections of the
leading arm are very much farther away ($\sim 50$ kpc) --- out of
range of our accessible instrumentation and requiring 10-m
telescopes should we ever desire to ``fill the gap" of our
coverage of the leading arm in the same way, and (4) by focusing
on rather nearby Sgr stars there is some potential for sample
contamination by Milky Way disk M giants.  We revisit the latter
possibility in \S 5.

To facilitate our discrimination of Sgr stream targets from other
Milky Way stars we take advantage of the ongoing studies of M
giants in the stream that are the focus of this series of  papers.
Apart from their intrinsic luminosity, M giants confer a
particular advantage in the study of the Sgr stream in that, as
Paper I demonstrated, the Sgr stream has contributed the majority
of the M giants found in the Milky Way halo. Thus, M giants
selected far enough away from the disk already have a high
likelihood of being from Sgr.\footnote{In addition, as was shown
in Paper I, using combinations of 2MASS colors it is possible to
cleanly separate M giants from any potential nearby, contaminating
M dwarfs --- though these should be fairly rare.} Figure 1
(adapted from Fig.\ 9 of Paper I) shows the distribution of M
giants with $(J-K_s)_o > 1.00$ lying within 10$^{\circ}$ of the
nearly polar Sgr orbital plane, as derived in Paper I.  Stellar
distances from the Sun (at the center) in this representation are
given by the corresponding dereddened $K_{s,o}$ magnitudes.  This
kind of map has the benefit of creating an approximate relative
spatial distribution free of biases imposed by presuming
particular metallicities and color-magnitude relations needed to
convert apparent magnitudes to photometric parallax distances, and
works best when stars of a limited color range are
used.\footnote{See similar representations using stars with colors
filtered to be at the main sequence turn-off in Newberg et al.
(2002), for example.} Since most of the M giants in the figure lie
in the range $1.0 < (J-K_s)_o < 1.1$, this magnitude-based
distribution reveals the basic structure of the Milky Way and Sgr
stream (modulo metallicity-based variations in the absolute
magnitudes of these stars), albeit with an approximately
logarithmic distance scale. This log scale has the benefit not
only of compressing the apparent width of the distant parts of
both the Sgr leading and trailing arm, making them more visible,
but of expanding the relatively small volume of space occupied by
stars we have targeted in the northern Galactic hemisphere, to
make their relative positions more clear.  However, as pointed out
in Paper I, the substantial stretching of the nearby Sgr leading
arm in such a rendition makes it appear more diffuse than it
really is.  The reader is directed to Figures 10 and 11 in Paper I
for a linear distance version of this distribution where the
nearby leading arm is less ``fuzzed out", and to Figure 9 of that
paper for a ``clean version" (without colored dots) of the Figure
1 distribution, for comparison.  The reader is also referred to
Figure 1 of Paper IV for an N-body model representation of the
observed debris that provides a useful guide to the expected
positions of leading (and trailing) arm debris in the Sgr orbital
plane.

Figure 1 (and its modeled counterpart in Paper IV) provides one
basis on which stars were selected for study here. But, in
addition to specifically targeting M giant stars apparently {\it
positioned} in particular portions of the Sgr leading arm, we also
pre-select stars that have radial velocities appropriate to these
positions based on Sgr debris models (Figure 10 of Paper IV)
constrained to fit all available positional and radial velocity
data for Sgr (e.g., Fig.\ 2).  The velocities used for this
project --- both those of the stars we targeted here and those
that provide the constraints for the fitted models --- have been
collected through an ongoing medium resolution spectroscopic
survey of 2MASS M giants (Paper II, Majewski et al. in
preparation; see also the data presented in Paper
IV).\footnote{However, the echelle spectra obtained here allow us
to derive improved velocities, and these new velocities are also
presented in Table 1 (see \S3.2).} Figure 2 shows, as a function
of the Sgr orbital plane longitude ($\Lambda_{\sun}$), the
observed radial velocities, converted to the Galactic Standard of
Rest (GSR), of M giants lying near the Sgr orbital plane.  The
rather velocity-coherent trend of the Sgr trailing arm (not
explored here) is obvious on the right.  The RV distribution of
leading arm stars is less coherent, especially where it comes
close to the Sun, because of the considerable angular spread of
the stream on the sky at this point (and therefore a wider
variation in the projection of the stellar space motions on the
line of sight). Additional RV spreading in the leading arm occurs
because of the greater overlap of stars with different orbital
energies at the same orbital phase compared to the trailing arm
(See Fig.\ 1 of Paper IV). The trend of Sgr leading arm stars in
Figure 2 is sinusoidal (see also Fig.\ 10 of Paper IV).  From left
to right in Figure 2: (1) Leading arm stars are first moving away
from the Sgr core (at $\Lambda_{\sun} = 0^{\circ} = 360^{\circ}$)
and have positive $v_{GSR}$ at high $\Lambda_{\sun}$; (2) after
apo-Galacticon the leading arm bends towards the general direction
of the Sun, and leading arm stars develop negative $v_{GSR}$ which
continue to decrease as the leading arm curves towards the solar
neighborhood and approaches from the general direction of the
North Galactic Cap (NGC, centered near
$\Lambda_{\sun}=256^{\circ}$); (3) as the leading arm traverses
the Galactic plane near the Sun, the $v_{GSR}$ changes sign again
with the trailing arm stars now speeding away from the solar
neighborhood and arcing under the Galactic Center ($\Lambda <
100^{\circ}$).  It is worth noting that after passing below the
Galactic plane, the leading arm crosses the trailing arm; the
velocity trends of the two arms also cross in this region
($\Lambda < 100^{\circ}$) as shown in Fig.\ 10 of Paper IV.
Because the leading arm has yet another apogalacticon at $\Lambda
< 100^{\circ}$, the debris, and the associated velocities, is
expected to become less coherent.  This can be seen by the green
points to the lower right of Figure 1 in Paper IV, but is not
obvious by Figure 10 of that same paper, which did not show this
dynamically older debris.  That the overall spatial and velocity
distribution of the leading arm at this point becomes more diffuse
can also be seen in the models of Ibata et al. (2001; see their
Fig.\ 3).

\subsection{Spectroscopic Observations}

Figures 1 and 2, and the associated figures from our models in
Paper IV, guided the selection of four samples of stars for
analysis here:

(1) A large sample of stars (red symbols in Figs.\
1 and 2) were selected to have both positions and velocities
consistent with being in the leading arm north of the Galactic
plane, and in the general direction of the NGC (with Sgr
longitudes $\Lambda_{\sun}$ = 220-290$\arcdeg$). Of these, 21 were
observed with the $R$=$35,000$ resolution Mayall 4-m Echelle on
the nights of UT 05-09 May 2004.  On UT 10-13 Mar 2004,
$R$=$46,000$ SARG spectra for nine additional M giants in the same
part of the stream were obtained with the TNG telescope in the
Canary Islands.

This ``leading arm" sample is the largest in our survey, because of
our mostly northern hemisphere telescope access.  A large range of
$K_{s,o}$ has been explored, partly because when weather conditions
were non-ideal we resorted to brighter, generally closer stars.  Indeed, some
of the stars explored have initially projected (i.e. Paper I) distances as low
as 1 kpc.  Stars this close do lie among the Galactic thick disk stars, but
when selecting such stars we deliberately chose stars that lie along the
leading arm trend in Figure 2, and which, for the most part, have strongly
negative $v_{GSR}$'s (e.g., $<-65$ km s$^{-1}$) that are unlike the typical
thick disk star.

Nevertheless, as a means to explore and limit the extent to which
our analysis of this leading arm sample may have been affected by
thick disk contaminants that just happen to have the ``right"
velocity, we further divide this group even into a ``best"
subsample (the fainter, generally farther seventeen stars that are
very highly likely to be in the Sgr leading arm) and a ``less
certain" subsample of thirteen stars, including those stars marked
with red symbols within the boundary drawn in Figure 1.  The
latter subsample includes the ten leading arm north stars with
$K_{s,o}<7.5$ as well as three stars at the highest
$\Lambda_{\odot}$ that are closer to the Galactic bulge. If there
is contamination of the leading arm north group by thick disk
stars, it will most likely be among the latter sample, which has
initially estimated distances from 1-5 kpc (based on the
color-magnitude relation for an [Fe/H]$\sim$-0.4 population
assumed in Paper I).\footnote{We will show below
that these distances are, in the mean, underestimated because most
of the stars are more metal-poor than [Fe/H]=-0.4.}  We further
discuss the issue of contamination, and the fact it is not
expected to be affecting the overall conclusions of this study, in
\S 5.

(2) Ten M giant stars with positions and velocities of leading arm
stars south of the Galactic plane (green symbols) were observed
with the $R$=$19,000$ MIKE spectrograph on the 6.5-m Clay
telescope at Las Campanas Observatory on the night of UT 15 Aug
2005. These stars, with $\Lambda_{\sun}$ = 20-45$\arcdeg$, include
stars with projected distances both inside and outside of the
trailing arm and with $v_{GSR}$ well away from the trailing arm
trend (Fig.\ 2). According to the models of Paper IV, the leading
arm stars south of the Sun were predominantly stripped from Sgr
roughly 2-3 Gyr ago whereas those now north of the Sun were
stripped roughly 1.5-2 Gyr ago.

(3) Six stars in the very center of the Sgr core (magenta symbols)
were also observed with MIKE on the same observing run as the
other southern Sgr stars.  Unlike the other groups of stars we
looked at in this survey, these Sgr core stars were not pre-vetted
based on radial velocity data, but rather selected on the basis of
the infrared color-magnitude diagram. Based on the high density of
Sgr giants in the core, this was a relatively safe strategy.  We
subsequently derived radial velocities for these stars from the
MIKE spectra (values shown in Table 1), and these show them all to
have radial velocities consistent with the Sgr core.  These
velocities were obtained via cross-correlation against four radial
velocity standards using the echelle order we used for the stellar
atmospheres analysis described in \S4.

We combine this small sample of Sgr core stars with the other extant
echelle resolution metallicities for Sgr core stars in the literature in our
analysis of the MDF below.

(4) Finally, we targeted thirteen additional M giants (blue
symbols) lying among the stars of the Sgr leading arm in the NGC
that were found to have velocities quite unlike that expected for
the Sgr leading arm at this position.  We refer to this sample
as the ``North Galactic Cap (NGC) group".  Most of these stars are too
far away and have velocities far too large to be contamination
from the Galactic disk.  On the other hand, while dynamically old
Sgr stars from the wrapped {\it trailing arm} --- if they exist in
the M giant sample --- are expected to lie in the direction of the
NGC (Fig.\ 1 of Paper IV) and with more positive radial velocities,
initial estimates of the distances of the NGC group stars from
the Paper I photometric parallax analysis (which, again, assumes an
[Fe/H]$\sim$-0.4 giant branch color-magnitude relation) puts these stars {\it
too close} to the Sun to be consistent with wrapped trailing arm
debris. Thus, obtaining echelle resolution spectra of some of
these peculiar stars is of interest in order to test whether they
can be ``chemically fingerprinted" as Sgr debris (\S 6 and 7).

To lessen potential metallicity biases, M giant stars in all four
groups were selected with a wide range of $J-K_s$ color ---
typically $\sim$1.0-1.2. Otherwise, the specific selection of
targets was dictated by the desire to sample the four groups of
stars outlined above and by the limitations of assigned observing
schedules. Table 1 summarizes the targets, their equatorial and
Galactic coordinate positions, dereddened 2MASS $K_s$ and $J-K_s$
photometry from Paper I, the Sgr orbital plane longitude
($\Lambda_{\sun}$), the velocity in the Galactic Standard of
Reference ($v_{GSR}$), and the spectrograph with which each target
was observed and on what date. For most stars in Table 1 we give
two velocities:  The first is from the medium resolution
spectroscopic campaign described above (\S3.1), which has typical
velocity uncertainties of about 5-15 km s$^{-1}$; these are the
velocities that were used in the selection of the present
spectroscopic samples and that are shown in Figure 2. The second
$v_{GSR}$ values were derived from the new echelle resolution
spectra by cross-correlating the echelle order that we use for the
chemical analyses (presented below) against that same order for
several radial velocity standard stars taken from the Astronomical
Almanac. The estimated velocity errors for the echelle data are
1.6 km s$^{-1}$ for the MIKE spectra, 0.6 km s$^{-1}$ for the KPNO
spectra, and 0.2 km s$^{-1}$ for the SARG spectra. As may be seen,
the echelle and medium resolution velocities track each other
well, with a dispersion in their difference of 7.3 km s$^{-1}$,
which is consistent with the uncertainties in the medium
resolution spectra. In the case of the Sgr core stars we only have
velocities derived from the new, echelle spectra. Table 1 also
gives the $S/N$ of each spectrum; these ranged from $\sim$40-190
for the Mayall, $\sim$110-390 for the TNG and $\sim$35-120 for the
MIKE data. The $S/N$ was determined using the total photoelectron
count level at 7490\AA.

\section{Iron Abundance Analysis}

\subsection{Data Reduction and Equivalent Width Measurements}

To convert our 2-D echelle images into fully calibrated 1-D
spectra we used the basic echelle spectra reduction routines in
the Image Reduction and Analysis Facility (IRAF).\footnote{IRAF is
distributed by the National Optical Astronomy Observatories.} This
process included overscan and bias correction, scattered light
subtraction, flattening of the spectra by division of normalized
quartz lamp exposures, extraction of the echelle orders,
wavelength calibration using paired exposures of either a thorium
(SARG spectra) or a thorium-argon discharge tube (KPNO and MIKE
spectra) taken at the same telescope position as each target
observation, and spectrum continuum fitting.

For the present analysis we focused on eleven unblended
\ion{Fe}{+1} lines (listed in Table 2) found in a particular part
of the spectrum previously explored by Smith \& Lambert (1985;
1986; 1990
---hereafter ``S\&L") in their spectroscopic exploration of M
giants (see Section 4.3). We used the IRAF task splot to measure
interactively the equivalent widths (EWs) of these lines, which
typically spanned one echelle order.

Because three different instruments (with three different
resolutions --- see examples of spectra from each instrument in
Fig.\ 3) were used to collect the spectra, the possibility that
the equivalent widths might suffer from significant systematic
differences was investigated.
In Figure 4 we compare the measured EWs of \ion{Fe}{+1}
lines in very high $S/N$ spectra of Arcturus (the one star we
have observed on all three systems) taken on each the SARG, KPNO and MIKE
spectrographs.  The equivalent widths
for the three different spectrographs agree reasonably well.
Only slight offsets of EW(Mayall)$-$EW(SARG)=$11.0\pm10.7$ m\AA\  and
EW(MIKE)$-$EW(SARG)= $4.9\pm3.8$ m\AA\  were found; because of the
sizes of the uncertainties on these offsets compared to their
measured values, we elected not to apply any corrections
between spectrographs.  However, if real, the level of these offsets in terms
of an [Fe/H] value is +0.09 dex and +0.04 dex, respectively, offsets about
those size of the estimated random [Fe/H] errors (see below).

The final measured EWs of the \ion{Fe}{+1} lines for each of the
Sgr spectra are given in Table 3.  We also include there
the EW's measured for Arcturus from spectra taken on the the three
different instruments used to make Figure 4, as well
as for several standard stars we analyze next.

\subsection{Determining the Effective Temperatures, Surface Gravities, and Iron Abundances}

A detailed abundance analysis from spectra requires as input
parameters the stellar effective temperature, $T_{\rm eff}$,
surface gravity (usually parameterized as $\log g$), and
metallicity. The first parameter, $T_{\rm eff}$, has been
determined using the dereddened 2MASS ($J-K_{\rm s}$) colors and
the Houdashelt et al. (2000) color-temperature
calibration.\footnote{Houdashelt et al. (2000) work in the CIT
near-infrared filter system, whereas our Sgr star photometry is in
the 2MASS system.  We adopted the Carpenter (2001) transformation
equations to convert the 2MASS colors to the CIT system.} In the
following analysis, the effective temperature is used in
combination with stellar isochrones (Girardi et al. 2000; Demarque
et al. 2004, hereafter Y$^2$) to constrain the stellar surface
gravity.

For a given population age and metallicity, a single
isochrone defines a nearly unique curve in a
$T_{\rm eff}$-$\log g$ plane, so that a given effective
temperature defines a $\log g$ value.  Red giants can
either be first ascent red giant branch
(RGB) stars or asymptotic giant branch (AGB) stars and these
two separate phases of stellar evolution define
slightly different $T_{\rm eff}$-$\log g$ tracks.  However, the
$\log g$ differences for a given $T_{\rm eff}$ are quite
small in older stellar populations.  This
is particularly true for red giants with M-star
temperatures ($T_{\rm eff}$$\le$ 4000K), where the
RGB and AGB almost coincide in the $T_{\rm eff}$-$\log g$
diagram (and where differences between the RGB and AGB are
measured in hundredths of a dex in $\log g$).

In principle then, the effective temperature in an old red giant
defines its $\log g$. The two other primary
variables that define the $T_{\rm eff}$--$\log g$ curve are age and
metallicity.  All of the potential Sgr populations are ``old'',
which here translates to ages greater than about 3 Gyr.  For a
specific metallicity, the difference between a 3 Gyr and a 10 Gyr
isochrone in a $T_{\rm eff}$-$\log g$ plane is not large (about 0.1
dex in $\log g$ at $T_{\rm eff}$=3800 K).  This is due to the small
difference in mass between a 3 Gyr red giant
($M\sim1.4$ M$_{\odot}$) and a 10 Gyr one ($M\sim1.0$ M$_{\odot}$).
Once a population is older than a few Gyr, the exact age becomes
relatively unimportant in defining $\log g$. Metallicity, on the other
hand, does have a significant effect on the derived $\log g$ for a
given effective temperature in an old red giant.  This effect is
incorporated into the abundance analysis here via an iterative
scheme matching the isochrone used to define $\log g$ to the iron
abundance then derived with that particular isochrone.  Sample
Fe I lines are used along with the photometric $T_{\rm eff}$ and an
initial estimate of $\log g$
from an isochrone of a given metallicity to derive [Fe/H].  If
this value of [Fe/H] does not match the adopted isochrone metallicity,
a new isochrone is selected and
the process is repeated until there is agreement between isochrone
and derived spectroscopic stellar metallicity.

The Fe I lines used to determine the iron abundance and final
isochrone metallicity (and thus the final $\log g$) are listed in
Table 2, along with the excitation potentials and $gf$-values. The
Fe I $gf$-values in Table 2 were determined by measuring these Fe
I equivalent widths in the solar flux atlas of Kurucz et al.
(1984) and varying the $gf$-values for each line in order to match
the solar iron abundance of $A$(Fe)=7.45 (Asplund, Grevesse, \&
Sauval 2005).  The analysis here used the LTE code MOOG (Sneden
1973) combined with a Kurucz ATLAS9 (1994) solar model, with
$T_{\rm eff}$=5777 K, $\log g$= 4.438, and a microturbulent
velocity, $\xi$=1.0 km s$^{-1}$.

A comparison of the Fe I $gf$-values derived in this way
with those given for these same lines in
Kurucz (1995) line list yields a difference of $\Delta \log gf$=
+0.14$\pm$0.15.  This is a small offset between these two
$gf$-value scales, with a small dispersion comparable
to the measured line-to-line variations found when the program
stars were analyzed.

The model atmospheres adopted in the analysis were generated by
interpolation from the Kurucz (1994) grids.\footnote{From
http://kurucz.harvard.edu/grids.html.}
In our iterative scheme, we also must
assume an initial metallicity for the model atmosphere.
Both this and the isochrone used to estimate $\log{g}$ are
iterated until the derived iron abundance of the stars
agrees with the metallicity of the model atmosphere, and
the metallicity of the adopted isochrone.

\subsection{An Analysis of Nearby ``Standard'' M Giants}

The abundance analysis method described in the previous section
can be tested on nearby, well-studied M giants that have physical
properties that bracket approximately those of the program Sgr
stream red giants. Included in the observed dataset for this
program are three nearby M giants ($\beta$ And, $\rho$ Per, and
$\beta$ Peg) that were analyzed in a series of papers by S\&L.
S\&L focussed their studies on a narrow spectral window, near
$\lambda$7440-7590\AA\, for abundance determinations in M, MS, and
S stars.  This region is quite free from significant TiO
blanketing down to temperatures of about $T_{\rm eff}$=3200-3300K
in giant stars, which allows for a straightforward abundance
analysis.  Smith \& Lambert exploited this fact to explore
nucleosynthesis in cool red giants on both the RGB and AGB.  The
same spectral region is used in this study for the Sgr stream M
giants and the three bright M giants that were analyzed by S\&L
are analyzed here using the techniques described  in Section 4.2.
Along with $\beta$ And, $\rho$ Per, and $\beta$ Peg standard stars
we include $\alpha$ Tau, the K5III giant used by S\&L as their
standard star.

As a first comparison of the spectra collected here with those
from S\&L, eleven Fe I lines, common to both studies, were
measured in the three M giants and the mean difference in
equivalent widths is found to be EW(this study)$-$EW(S\&L)
$=-6\pm$7 m\AA. This small offset is not significant and the
scatter is about what is expected given the overall
signal-to-noise levels and spectral dispersions.  Spectra from
this study and those from S\&L are of comparable $S/N$ and
resolution and have expected equivalent-width uncertainties of
about $\pm$5m\AA.  Differences between the two sets of
measurements would then be expected to scatter around
5$\times$(2)$^{1/2}$ or $\pm$7m\AA --- i.e., close to what is
found.

Stellar parameters were derived for $\alpha$ Tau, $\beta$ And,
$\rho$ Per, and $\beta$ Peg using first a method similar to that
used by S\&L, followed by the method used here for the Sgr stream
stars (\S 4.2) to see how these different methods compare in
deriving $T_{\rm eff}$, $\log g$, and [Fe/H].  S\&L used ($V-K$)
colors to define $T_{\rm eff}$, while they set the luminosity
based on the Wilson (1976) calibration of the strength of the Ca
II K-line with absolute visual magnitude ($M_{\rm V}$).  Given
luminosity and effective temperature, S\&L then compared these
observed values to stellar-model mass tracks to set $\log g$ via
the relation of $g \propto (M \times L)/T_{\rm eff}$$^{4}$.

One significant difference between this particular S\&L procedure
and our modified use of it here concerns the estimate of the
luminosities. The S\&L studies predate the availability of
Hipparcos parallaxes, which are now well-measured for the four red
giants under consideration. Table 4 lists the Hipparcos parallaxes
for $\alpha$ Tau, $\beta$ And, $\rho$ Per, and $\beta$ Peg, as
well as the resulting distances (and their respective
uncertainties).  These distances then provide the absolute $V$-
and $K$-magnitudes also listed in the table (with the distance
uncertainties considered).  Both $V$ and $K$ bolometric
corrections were applied to determine $M_{\rm bol}$ in Table 4,
with the respective corrections differing by less than 0.05 in
magnitude. Finally, effective temperatures from both a ($V-K$)
calibration (Bessell et al. 1998) and the ($J-K$) calibration from
Houdashelt et al. (2000) are listed in Table 4.\footnote{In this
case, the near infrared colors for the bright stars are in the
Johnson system, and we converted to the Houdashelt et al. (2000)
CIT system using the transformation equations in Bessell \& Brett
(1988).}

Stellar luminosities for the four standard red giants are
calculated by adopting $M_{\rm bol}$=4.74 for the Sun and the values
of $\log$($L$/L$_{\odot}$) versus the mean $T_{\rm eff}$ (i.e.
the average of the two determinations in the previous paragraph)
are plotted in the
two panels of Figure 5. Also plotted in this figure are stellar
model tracks from the Padua grid
\footnote{http://pleiadi.pd.astro.it} for masses of $M=$1.0, 1.5,
and 2.0M$_{\odot}$.  The top panel shows models with near-solar
metallicity ($Z=$0.019), while the bottom panel has models with
[M/H]$\sim=-0.4$ ($Z$=0.008). This figure illustrates the effect
that metallicity has on estimates of the gravity.  At lower
metallicities the model tracks indicate a lower mass for a given
measured $T_{\rm eff}$ and $\log L$. This effect is quantified in
Table 5 where $T_{\rm eff}$ and $\log L$/L$_{\odot}$ are listed,
along with the estimated mass and resultant $\log g$ for the two
model metallicities plotted in Figure 5.

Given the effective temperatures and model mass (and thus $\log
g$) as a function of metallicity, the \ion{Fe}{+1}
equivalent-widths are used in an abundance analysis to achieve
final agreement between derived [Fe/H] and model metallicity.  In
the line analysis the microturbulent velocity is set by the
requirement that the derived Fe abundance be independent of the Fe
I equivalent width for the different lines. The derived values of
$\log g$, microturbulence ($\xi$) and [Fe/H] are listed in Table
5.  These values of $\log g$ can be referred to as ``Hipparcos
gravities'' because they are set by the mass, which is derived
from the luminosity, which is derived from the distance, which is
derived from the Hipparcos parallaxes. This analysis is very
similar as that used by S\&L, differing only in that S\&L used
\ion{Ca}{+2} K-line absolute magnitudes to establish a luminosity
while here the Hipparcos parallaxes are used to get a distance and
therefore a luminosity.

With the basic red giant parameters now defined for the bright
giant stars via the standard Fe-abundance analysis, the new
analysis technique (\S4.2) used in this paper for the candidate
Sgr stream red giants can be checked for differences when also
applied to these same bright giant stars.  Recall that with Sgr
stream stars there is no reliable distance estimate available to
establish luminosity; rather, the effective temperature is used in
combination with the Fe abundance to establish surface gravity via
isochrone tracks.  Moreover, for the new analysis the $T_{\rm
eff}$ are derived only from ($J-K$) colors (rather than from both
$J-K$ and $V-K$ colors) due to the larger effects of uncertain
reddening on optical colors and also the fact that we don't have
$V-K$ colors for the Sgr stream giants. Finally, we apply several
different isochrone ages as well as two separate families of
isochrones (Girardi et al. 2000 versus $Y^2$) in the
characterization of the standard red giants to test the
sensitivity of the new technique to these variables.  The results
of the ``new" analysis applied to the bright giants are tabulated
in Table 6.

The Table 6 results show, first, that there is rather little
difference in the derived surface gravities to either the adopted
set of isochrones or the variation from 1.0 Gyr to 2.5 Gyr
isochrones. We have already mentioned (\S4.2) that there is only a
$\log g$ difference of 0.1 between a 3 and a 10 Gyr isochrone of
the same metalliticity; the 1 and 2.5 yr isochrones here are
intended to explore ages more appropriate to disk-like giants like
our standard stars, but we note that there is only a $\Delta
\log{g}$ difference of 0.05 between a 2.5 and a 5 Gyr isochrone of
the same metallicity. Moreover, a comparison between the Table 6
gravities and abundances and those derived from the more standard
analysis reveals no large differences.  Figure 6 provides a
graphical comparison of the surface gravities (top panel) and
[Fe/H] (bottom panel) derived from the two techniques, and shows
their close correlation.

This comparative analysis of the four red
giants with well-established, fundamental stellar parameters
demonstrates that the analysis technique used for the candidate
Sgr Stream red giants is sound, and yields reliable stellar
parameters and Fe abundances.

\subsection{Final Results}

Table 7 gives the results of the \S4.2 abundance analysis applied
to our Sgr stars.  For each star, the columns give the derived effective temperature
using the Houdashelt et al. (2000) color-temperature relation applied to
the 2MASS $(J-K_s)_o$ color, and the derived values of
the surface gravity ($\log{g}$), microturbulence and [Fe/H].  In the case
of the surface gravities, any entry given as ``0.0(-)" means that our
iterative procedure was converging on a model atmosphere with
$\log{g} < 0$,
whereas the Kurucz (1994) model atmosphere grids do not go below
$\log{g} = 0$.  In these cases, we have adopted the $\log{g} = 0$
atmosphere.

The final column in Table 7 represents the standard deviation in
the line abundance determinations. In principle, from the adopted
model atmosphere and each EW we get a measure of the abundance.
With multiple EWs from different Fe I lines, MOOG calculates the
standard deviation of the resulting abundances. The typical
standard deviations are about 0.1 dex.  Combined with the
instrument-to-instrument offsets discussed in \S4.1 and shown in
Figure 4 as well as other potential offsets, such as those shown
in Figure 6, we estimate the full [Fe/H] errors, systematic and
random combined, to be no more than $\sim0.2$ dex.

\section{Metallicity Distribution Functions}

\subsection{The Sagittarius Core}

Figures 7 and 8 summarize the MDFs determined for the three groups
of Sgr core/leading arm samples studied here (Figure 7 shows the
distributions with the same absolute vertical scale, Figure 8
shows the distributions with the same normalized, fractional MDF
scale in each panel).

For the Sgr core, data for our six stars (Figs.\ 7a and 8a; shown
by the open histogram) have been combined with previous echelle
data for 14 K giants by Smecker-Hane \& McWilliam (2002) and for
15 M giants by Monaco et al.\ (2005). The precisions in the
metallicities quoted for each of these studies is 0.07 and 0.20
dex, respectively, similar to our results here. The combined MDF
from these data shows the very broad distribution previously
reported for Sgr (see \S2), with a peak near [Fe/H]=-0.3 but a
very long, metal-weak tail. The new MIKE spectra we collected
contribute two stars near [Fe/H] = -1 but the other four lie in
the metal-rich end of the distribution, and include one star we
determine to have solar [Fe/H]. We consider this star to be a bona
fide member of Sgr because of its chemical peculiarities (in
particular, its Ti, Y, and La abundances, which are like other Sgr
stars of similar metallicity, as we shall show elsewhere --- Chou
et al., in preparation).

\subsection{Leading Arm North}

Panels (b) in Figures 7 and 8 present the MDF for all stars we
selected to be members of the Sgr leading arm in the Northern
Hemisphere. As may be seen, while broad like the MDF of the Sgr
core, the distribution ``of leading arm north" stars is, on the
whole, more metal poor than the Sgr core, with a median near -0.7
dex.

As discussed in \S3, this particular sample is
the most vulnerable to potential contamination by Milky Way disk M giants.
However, several arguments can be made that this contamination is probably
small, and, even if there is some contamination, it has little affect on the overall
conclusions of the present study:

(1) First, we can compare the MDFs of subsamples of ``leading arm
north" stars, divided into the ``best" (generally farther) and
``less certain" (generally closer) Sgr stream groups discussed in
\S3.  Figure 9 makes this comparison, and shows that there is
little difference in the overall character of the two MDFs.  The
two subsamples have the same median [Fe/H] and similar tails to the
metal-rich end.  The difference in the mean metallicities of the
two samples, -0.72 and -0.64 dex, respectively, is much smaller
than the MDF dispersions (0.31 and 0.33 dex, respectively).

(2) The majority of the stars in the Leading Arm North sample are more metal-poor
than the mean metallicity of the Sgr core, so that their projected distances are
even farther away from the Milky Way disk than initially projected based on
the Paper I photometric parallaxes that assumed a Sgr core RGB color-magnitude relation.
For the ``best" subsample, the implied minimum distances are generally 10 kpc or more,
well above the Galactic disk.

(3) The median metallicity of the Galactic thick disk, the Milky Way component most
likely to contribute contaminants, is well known to be about -0.7 dex (whereas
the thin disk would contribute more metal rich stars in general, if at all).  Thus, we might expect
the probability distribution of Milky Way contaminants to look very similar to the
distribution we actually see, and therefore have little impact on the true MDF.

(4) As we shall show elsewhere (Chou et al., in preparation), the
abundance patterns (e.g., the combinations of [Fe/H], [Ti/Fe],
[Y/Fe], [La/Y]) of all but a few of the stars in the leading arm
north sample (and indeed in our entire survey) are quite unlike
those of Milky Way stars, but very much resemble the patterns seen
in dSph stars, including Sgr (Bonifacio et al.\ 2000; Fulbright
2002; Smecker-Hane \& McWilliam 2002; Shetrone et al.\ 2003; Venn
et al.\ 2004; Geisler et al.\ 2005; Monaco et al.\ 2005).

(5) The leading arm stars were
pre-selected to be in the Sgr stream and to follow the expected
velocity trends for Sgr debris.  No evidence for other
M giant tidal debris from any other satellite is found to
intersect the Sgr stream.\footnote{While the Monoceros stream {\it
does} also contain M giant stars (Rocha-Pinto et al. 2003, Crane
et al. 2003), these lie outside of the Galactic disk along the
Galactic plane and not near the samples we have selected here. We
shall show in \S 6 that the NGC moving group M giants are also
likely to be from Sgr.}  Because the bulk of the
halo M giants are found to be contributed from the Sgr system and we are
probing the general orbital plane of the Sgr system and well away from the Galactic
disk for the most part, it is logical to
conclude that our leading arm samples (both north and south)
are indeed dominated by members from the Sgr dSph.

Thus, we expect the relative contamination of our leading arm north sample by Milky Way
stars to be small.  While at this point it is true that we cannot be assured
that {\it every star} in any of samples, or any one particular star within them,
is definitely a member of the Sgr stream, a few contaminants will have little effect on the
general conclusions of this paper, which are based on {\it mean trends} in the
Sgr MDF.  In this regard, it is sufficient that {\it most} of the stars are Sgr stream members
and to recognize that the Leading Arm North MDF differs significantly from that
of the Sgr core.

\subsection{Leading Arm South}

The Leading Arm South sample (Figures 7c and 8c) shows an even
more metal-poor MDF than either the Sgr core of the Leading Arm
North samples.   With regard to contamination by the Milky Way
disk, things are even more secure for this sample than for the
Leading Arm North: Not only are these stars even farther away from
the disk according to the original projected distances from Paper
I (and even more so if their projected distances are corrected
for their newly discovered low metallicity), but they have an MDF
even more unlike the Milky Way disk.  The median metallicity of
around -1.1 dex, the lack of stars more metal rich than
[Fe/H]=-0.7, the relatively small [Fe/H] dispersion in this
sample, and unusual chemical abundance patterns found in these
stars (Chou et al., in preparation) all argue against the notion
of significant contamination of this group of stars by the thick
disk.

\subsection{Evolution in the Sagittarius MDF}

Comparison of the Sgr core MDF with those at the two points in its
leading arm we explored here (Figs.\ 7b/8b and 7c/8c) reveals
substantial evolution in the Sgr MDF with position.  While all
three points of the Sgr system sampled contain stars from a
metal-poor population with [Fe/H] $<-1$, the relative proportion
of these stars increases with separation from the Sgr core.  The
latter shows a dominant metal-rich population peaked at [Fe/H]
$\sim-0.3$, whereas the median metallicity declines from
$\sim-0.4$ dex in the core to $\sim-0.7$ dex in the leading arm
north of the Sun and $\sim-1.1$ dex south of the Sun, which
represents debris lost from the Sgr core some 3.5 orbits
($\sim2.5-3$ Gyr) ago (Paper IV).

While the Figure 7c/8c MDF has only one star with [Fe/H]$>-0.95$,
because we are color-selecting {\it M giants} our samples tend to
be biased against finding {\it metal-poor giants} (which are bluer
and earlier in spectral type). Thus the significant, $-0.7$ dex
median metallicity gradient shown in Figures 7 and 8 may actually
{\it underestimate} the true gradient of what already appears to
be a substantial MDF variation along the Sgr stream.\footnote{A
possible selection effect that would bias the survey in the
opposite direction might arise from the fact that metal-poor
giants tend to be brighter at a given color, and therefore
possibly more likely to be observed.  We believe that this is less
likely to be affecting our results based on the fact that there
are no significant differences between the MDFs of the two
subsamples of Leading Arm North stars divided primarily into two,
large apparent magnitude bins ($4.8\lesssim K_{s,o} \lesssim 7.5$
and $7.5 \lesssim  K_{s,o} \lesssim 9.7$) shown in Figure 9.} We
address the implications of this gradient in \S7.

\section{Evidence for Sgr Trailing Arm in the North}

In the course of our ongoing, medium resolution radial velocity
survey of Sgr M giants (e.g., Majewski et al.\ 2004) we identified
a subsample of M giants lying among leading arm stars at the NGC,
but having the {\it opposite} velocity expected for {\it falling}
leading arm debris there (see $v_{\rm gsr}>0$ black points near
$\Lambda_{\sun}$ = 260$\arcdeg$ in Fig.\ 12 of Paper IV). Because
of their {\it apparent} proximity to the Sun (solid blue points,
Figs.\ 1 and 2), the origin of these stars has been puzzling.
Thirteen of these peculiar velocity M giants with median
$\Lambda_{\sun}$=265$\arcdeg$ were targeted with the Mayall 4-m
and TNG SARG echelle spectrographs on the same observing runs and
to the same approximate $S/N$ as the NGC leading arm stars (\S3).

The relatively low metallicities of these $v_{\rm gsr}>0$ stars
(Figs.\ 7d/8d) indicates that the initial Paper I photometric
distances for these stars (based on an assumed [Fe/H]$\sim-0.4$;
Fig.\ 1) were underestimated by a mean factor of $\sim1.5$, based
on the color-magnitude sequences presented in Ivanov \& Borissova
(2002). Adjusting the distances for correct metallicities ---
minding the $v_{\rm gsr}$ of these stars and recognizing that the
models were not well constrained for {\it old} debris --- we find
reasonable consistency of these stars with the Sgr {\it trailing
arm} towards the NGC (see Fig.\ 12 of Paper IV).

Detailed abundance analysis supports this conclusion. The MDF of
these positive $v_{\rm gsr}$ stars (Figs.\ 7d/8d) fits the general
trend with Sgr mass loss epoch established by the leading arm data
(Figs.\ 7a-c/8a-c); as may be seen by comparing the mass loss
epoch sequences of the leading and trailing arms in the Paper IV
(colored point in Figure 1) model, stars in our leading arm south
sample and in the NGC sample, if it is indeed old trailing arm
debris, were torn from Sgr at approximately the same time. Thus it
is compelling that the MDFs in Figures 7c/8c and 7d/8d look very
similar to one another. In addition, this NGC moving group is
found to have similarly peculiar \ion{Ti}{0}, \ion{Y}{0} and
\ion{La}{0} abundance trends as stars in the Sgr leading arm (Chou
et al., in preparation), further supporting the idea of a common
origin with these latter stars.

If trailing arm stars are found toward the NGC it establishes with
certainty that the Sgr debris tracks at least 3 orbits (2.5-2.75
Gyr) of mass loss (Paper IV); because of much stronger phase
mixing of debris in the leading arm, this fact is not well
established by the apparent {\it length} of the Sgr leading arm
(although previous evidence that it may exist has been offered by
Martinez-Delgado et al. 2004).  Moreover, including the MDF in
Figures 7d/8d in the overall sequence shown in Figures 7a-c/8a-c,
lends further support to the overall notion that there is a
significant MDF variation along the Sgr stream.

\section{Discussion}

Because Sgr is reputed to have enriched to near solar metallicity
by at least a few Gyr ago (Lanfranchi \& Matteucci 2004;
Bellazzini et al. 2006; Siegel et al. 2007), the observed MDF
variation over the past 3.5 orbits (2.5-3 Gyr) of mass loss cannot
be due to an intrinsic variation of the instantaneous mean
metallicity of the Sgr system with time. Rather, it must point to
the shedding of successive layers within the satellite over which
there must have been an intrinsic MDF {\it gradient} (see also
Mart{\'i}nez-Delgado et al. 2004). However, the $> 0.7$ dex median
metallicity variation in the debris lost over a 2.5-3 Gyr
timescale is quite large and suggests the loss of stars over a
significant radius in the system.  For comparison, the strongest
[Fe/H] gradient observed in the Sculptor dSph is about 0.5 dex
over about $0\fdg2$ ($\sim 275$ pc), which is about 15\% the
apparent Sculptor tidal radius; however, this same 0.5 dex change
also represents the {\it entire} variation seen across the
$\sim75\%$ of the Sculptor tidal radius studied in detail so far
(Tolstoy et al.\ 2004). Sculptor seems to have among the strongest
net internal metallicity gradients among Milky Way dSphs (though
some M31 dSphs may have larger gradients; Harbeck et al.\ 2001);
for comparison, the now well-studied Carina dSph exhibits only a
$-0.2$ dex gradient from its core to its tidal radius (Koch et
al.\ 2006).  Moreover, no large metallicity gradient seems to
exist within the main body of Sgr now: Alard (2001) identified
only a $-0.2$ dex variation in mean metallicity from the Sgr core
to $7\fdg5$ down the major axis. While the position of the current
tidal radius in Sgr is still uncertain, Paper I argues that it is
likely to be only $\sim$ 3-4$\arcdeg$ (or Sgr would be too massive
to produce its observed dynamically cold tails); thus the Alard
observation likely pertains to the beginning of the metallicity
gradient {\it within the debris tail}.  Therefore, we must
conclude either (1) the destruction of Sgr over the past several
Gyr has been fine-tuned to mass shedding from a narrow progenitor
radial range over which there was an extraordinarily strong [Fe/H]
gradient for a dSph, or, (2) more likely, Sgr experienced a quite
rapid change in its binding energy over the past several Gyr,
which has decreased the tidal boundary of the satellite across a
broader radial range over which there would have still been a
large net metallicity variation, but a shallower and more typical
{\it gradient}.\footnote{Support for significant Sgr mass loss
over its past $\sim3$ orbits is that about half of the Sgr M
giants in the corresponding tails lie $30\arcdeg$ beyond the Sgr
center (Paper I).} Such a catastrophic change of state happening
so relatively recently (1/5 the Hubble time) points to a dramatic
event affecting Sgr's life several Gyr ago, perhaps a transition
to its current, destructive orbit.

Figures 7 and 8 not only provide the first direct evidence that
the satellites of today may {\it not} well represent the stars
they lost to the halo, but that this effect can be considerable.
If tidal mass loss is typical among other dSph systems, as seems
to be the case (e.g., Mu\~{n}oz et al.\ 2006a, 2007; Sohn et al.
2007), it might explain such puzzles as why: (1) the detailed
chemical abundances (e.g., [$\alpha$/Fe] vs. [Fe/H]) of satellites
today appear to differ from those observed in the halo field to
which they should contribute (e.g., Font et al.\ 2006), (2) a
system like the Carina dSph, which exhibits clear signs of tidal
disruption, presently holds a much larger fraction of
intermediate-age than old stars today (Majewski et al.\ 2000,
2002), and (3) there remains a G dwarf problem in dSph systems
(e.g., Koch et al.\ 2006; Helmi et al.\ 2006).  Such mass loss
shaping of the MDF prompts caution in attempting to interpret the
chemical evolution and star formation history of a dSph based on
stars left in its core (e.g., Tolstoy et al.\ 2003; Lanfranchi \&
Matteucci 2004).

To demonstrate this point, we approximate the total MDF of the Sgr
core several Gyr ($\sim$3.5 orbits) ago using two methods to
account for stars now in the tidal streams produced over that
time. In the first method (Fig.\ 10, blue lines), the normalized
MDFs in Figs.\ 8a-c represent their respective median {\it
Galactocentric} orbital longitudes and each leading arm star (as
identified in Fig.\ 11 of Paper I) is assigned a
longitude-interpolated version of these different MDFs. Regions
obscured by the Galactic plane or overlapping trailing arm are
``filled in" by reflecting the numbers of stars in the
corresponding part of the trailing arm as seen from the Galactic
Center (in the case of the first $50\arcdeg$ of leading arm) or by
extrapolating the observed stream density (for the farthest
175-300$\arcdeg$ of leading arm -- i.e. that part starting in the
solar neighborhood). In the second method (Fig.\ 10, red lines) we
use the Sgr disruption model for an oblate Milky Way halo from
Fig.\ 1 of Paper IV and assign the normalized MDFs in Figs.\ 8a, b
and c to leading arm model stars lost on the last 0.5 orbit (i.e.,
since last apogalacticon; yellow-colored debris in Fig.\ 1 of
Paper IV), 1.5-2.5 orbits ago (cyan-colored debris) and 2.5-3.5
orbits ago (green-colored debris) respectively, while for debris
lost 0.5-1.5 orbits ago (magenta-colored debris) we use the
average of Figures 8a and b. The model provides the relative
numbers of stars in each Sgr population (bound and unbound). Both
``Sgr-progenitor" MDFs generated are relatively flat, exhibiting a
much higher representation of metal-poor stars than presently in
the Sgr core. These regenerated MDFs are, of course, necessarily
schematic, because (1) The [Fe/H] spread of the net MDFs is, of
course, limited by the input MDFs, (2) an M giant-based survey is
biased {\it against} finding metal-poor stars, and (3) Sgr stars
with [Fe/H]$\sim -2$ have already been reported (see \S1;
ironically, the most metal poor stars shown in Fig.\ 7 are
contributed by the input MDF of the Sgr {\it core}, which includes
bluer giants as well as a larger overall sample of stars that
allows a higher chance of drawing stars from a low probability,
metal-poor wing in the distribution).  But Figure 10 illustrates
how critically the observed MDFs of satellite galaxies  may depend
on their mass loss/tidal stripping history.

We have discussed {\it integrated} MDFs as a function of position
in the Sgr system, but it is likely that, like other dwarf
galaxies, Sgr has had a variable star formation history including
possible ``bursts" (Layden \& Sarajedini 2000; Siegel et al.
2007), and that these produced populations with different, but
overlapping radial density profiles in the progenitor satellite.
The MDF gradients described here may relate more to differences in
the relative proportion of distinct populations than a smooth
variation in mean metallicity from a more continuous star
formation history. ``Distinct" Sgr populations are suggested by
the multiple peaks and general character of the Figure 7 MDFs (and
even more strongly by stream position variations of the abundances
of other elements, like lanthanum; Chou et al., in prep.). Earlier
suggestions of multiple Sgr populations include Alard (2001),
Dohm-Palmer et al.\ (2001), Smecker-Hane \& McWilliam (2002),
Bonifacio et al.\ (2004), and Monaco et al.\ (2005). Greater
resolution of the initial Sgr stellar populations, their former
radial distributions, and the Sgr enrichment history will come
from further scrutiny of its tidal debris, particularly along the
{\it trailing arm}. As shown in Figure 1 of Paper IV, leading arm
stars lost on different orbits (i.e., shed from different radial
``layers") significantly overlap in orbital phase position; this
``fuzzes out" the time (i.e. initial satellite radius) resolution.
In contrast, the dynamics of the longer trailing arm yields much
better energy sorting of the debris, and stars stripped at
specific epochs can be more cleanly isolated.  In addition, study
of the trailing arm will allow much better separation of the Sgr
debris from potential Milky Way disk M giant contamination.

The abundance gradients found here imply that the estimated
photometric distances for {\it many} M giant stars along the Sgr
tidal arms have been systematically underestimated in Paper I,
where photometric parallaxes were derived using the
color-magnitude relation of the Sgr core. The best-fitting Sgr
destruction models of Paper IV should now be refined to account
for this variation (as well as an updated distance for the Sgr
core itself --- e.g., Siegel et al. 2007). Proper spectroscopic
parallax distances will necessarily require assessment of both
[Fe/H] and [$\alpha$/Fe] to determine absolute magnitudes. We
undertake this task elsewhere.

\acknowledgements We gratefully acknowledge support by NSF grant
AST-0307851, NASA/JPL contract 1228235 and the David and Lucile
Packard Foundation as well as Frank Levinson through the
Peninsular Community Foundation. VVS and KC also thank support
from the NSF via grant AST-0307534 and AURA, Inc. through
GF-1006-00. D.G. gratefully acknowledges support from the Chilean
{\sl Centro de Astrof\'\i sica} FONDAP No. 15010003. The author
thanks Jon Fulbright for kindly providing the MIKE spectrum of Arcturus.
Parts of this paper were written while SRM, KC and VVS
participated in the ``Deconstructing the Local Group" workshop
held at the Aspen Center for Physics in June 2006.  Finally,
we appreciate helpful comments from the anonymous referee.

\begin{deluxetable}{cccccccccccr}
\rotate
\tabletypesize{\scriptsize} \tablecaption{The Program
Stars} \tablewidth{0pt} \tablehead{ \colhead{Star ID} &
 \colhead{$\alpha$(2000)} & \colhead{$\delta$(2000)} & \colhead{$l$}  &
\colhead{$b$} & \colhead{$K_{s,o}$} & \colhead{$(J-K_s)_o$} &
\colhead{$\Lambda_{\odot}$} & \colhead{$v_{\rm gsr}$(old/new)} &
\colhead{Spectrograph} & \colhead{Observation} & \colhead{$S/N$}
\\
\colhead{} & \colhead{(deg)} & \colhead{(deg)} & \colhead{(deg)} & \colhead{(deg)} &
\colhead{} & \colhead{} & \colhead{(deg)} & \colhead{(km s$^{-1}$)} & \colhead{} &
\colhead{UT Date} & \colhead{} } \startdata
\multicolumn{12}{c}{\it Sgr Core}  \\
$1849222-293217$ & 82.34253 & -29.53815 & 5.98090 & -12.58070 & 11.481 & 1.00 & 358.63837 & 135.9 & MIKE & 2005 Aug 15 & 34 \\
$1853333-320146$ & 283.38861 & -32.02935 &  4.00803 & -14.40648 & 11.240 & 1.05 & 359.97171 & 164.5 & MIKE & 2005 Aug 15 & 51 \\
$1854283-295740$ & 283.61789 & -29.96109 & 6.04514 & -13.76432 & 11.180 & 1.06 & 359.80359 & 162.3 & MIKE & 2005 Aug 15 & 43 \\
$1855341-302055$ & 283.89218 & -30.34867 & 5.77648 & -14.13644 & 11.392 & 1.03 & 0.10415 & 152.6 & MIKE & 2005 Aug 15 & 74 \\
$1855556-293316$ & 283.98166 & -29.55454 & 6.55899 & -13.89102 & 11.230 & 1.09 & 0.04467 & 173.9 & MIKE & 2005 Aug 15 & 45 \\
$1902135-313030$ & 285.55618 & -31.50829 & 5.24634 & -15.90276 & 11.198 & 1.06 & 1.70370 & 158.8 & MIKE & 2005 Aug 15 & 55 \\
\\
\multicolumn{12}{c}{\it Sgr North Leading Arm --- Best Subsample}\\
$0919216+202305$ & 139.83992 & 20.38467 & 208.89221 & 41.35083 & 8.663 & 1.09 & 212.41455 & -133.5/-125.4 & ECHLR & 2004 May 07 & 74 \\
$0925364+213807$ & 141.40163 & 21.63516 & 207.89902 & 43.12523 & 9.592 & 1.17 & 213.68213 & -239.4/-215.4 & ECHLR & 2004 May 07 & 54 \\
$1034395+245206$ & 158.66466 & 24.86820 & 209.33199 & 59.27583 & 9.140 & 1.11 & 228.44516 & -116.0/-102.3 & ECHLR & 2004 May 09 & 62 \\
$1100516+130216$ & 165.21519 & 13.03777 & 236.02568 & 60.56746 & 8.856 & 1.04 & 238.10635 & -194.1/-186.5 & ECHLR & 2004 May 06 & 77 \\
$1101112+191311$ & 165.29662 & 19.21981 & 224.41052 & 63.52243 & 9.146 & 1.07 & 236.06346 & -223.8/-219.2 & ECHLR & 2004 May 06 & 73 \\
$1114573-215126$ & 168.73872 & -21.85714 & 275.07312 & 35.74147 & 7.864 & 1.22 & 257.30670 & -193.0/-198.0 & ECHLR & 2004 May 06 & 72 \\
$1116118-333057$ & 169.04900 & -33.51587 & 281.07785 & 25.28218 & 7.697 & 1.15 & 266.35822 & -157.7/-140.6 & ECHLR & 2004 May 09 & 46 \\
$1140226-192500$ & 175.09427 & -19.41671 & 280.73941 & 40.37285 & 8.663 & 1.03 & 262.44983 & -204.1/-205.2 & ECHLR & 2004 May 06 & 80 \\
$1249078+084455$ & 192.28256 & 8.74870 & 301.12396 & 71.61227 & 9.295 & 1.05 & 264.23920 & -44.1/-53.6 & ECHLR & 2004 May 05 & 56 \\
$1318500+061112$ & 199.70825 & 6.18672 & 321.43869 & 68.06859 & 9.229 & 1.02 & 271.92865 & -24.7/-31.3 & ECHLR & 2004 May 06 & 70 \\
$1319368-000817$ & 199.90341 & -0.13814 & 318.02545 & 61.90507 & 7.741 & 1.22 & 275.23373 & -54.7/-41.6 & ECHLR & 2004 May 05 & 63 \\
$1330472-211847$ & 202.69652 & -21.31316 & 315.03806 & 40.63177 & 8.310 & 1.01 & 289.06110 & -183.8/-181.5 & ECHLR & 2004 May 09 & 41 \\
$1334532+042053$ & 203.72151 & 4.34796 & 329.30008 & 64.97141 & 9.598 & 1.08 & 276.31186 & 17.4/23.3 & ECHLR & 2004 May 06 & 57 \\
$1411221-061013$ & 212.84189 & -6.17026 & 336.00510 & 51.49488 & 9.512 & 1.08 & 289.51416 & -4.4/-6.7 & ECHLR & 2004 May 09 & 61 \\
$1450544+244357$ & 222.72687 & 24.73260 & 34.60439 & 63.09545 & 9.713 & 1.03 & 281.42078 & -59.5/-66.6 & ECHLR & 2004 May 06 & 61 \\
$1456137+151112$ & 224.05695 & 15.18672 & 16.93899 & 58.66600 & 7.122 & 1.05 & 288.11844 & 37.4/33.3 & SARG & 2004 Mar 11 & 128 \\
$1512142-075250$ & 228.05925 & -7.88056 & 352.23251 & 41.13720 & 9.531 & 1.11 & 303.44876 & 19.7/4.3 & ECHLR & 2004 May 05 & 65 \\
\multicolumn{12}{c}{\it Sgr North Leading Arm --- Less Certain Subsample}\\
$1111493+063915$ & 167.95526 & 6.65415 & 249.26958 & 58.71890 & 5.387 & 1.15 & 243.01091 & -96.4/-93.6 & SARG & 2004 Mar 11 & 367 \\
$1112480+013211$ & 168.19978 & 1.53646 & 256.03873 & 55.16073 & 5.673 & 1.04 & 245.24609 & -110.8/-135.4 & ECHLR & 2004 May 07 & 124 \\
$1128316-031647$ & 172.13158 & -3.27976 & 266.38379 & 53.60342 & 5.230 & 1.09 & 251.16019 & -98.1/-98.6 & SARG & 2004 Mar 11 & 376 \\
$1135388-022602$ & 173.91154 & -2.43394 & 268.24146 & 55.24917 & 5.825 & 1.10 & 252.52528 & -81.3/-78.4 & SARG & 2004 Mar 11 & 267 \\
$1208101-090753$ & 182.04225 & -9.13136 & 285.31238 & 52.25384 & 6.595 & 1.07 & 263.57907 & -84.8/-85.9 & SARG & 2004 Mar 11 & 169 \\
$1223590-073028$ & 185.99593 & -7.50770 & 291.09094 & 54.73156 & 4.820 & 1.16 & 266.45792 & -146.1/-144.3 & ECHLR & 2004 May 06 & 174 \\
$1224255-061852$ & 186.10632 & -6.31443 & 290.89117 & 55.92445 & 6.864 & 1.05 & 265.95511 & -64.8/-64.7 & SARG & 2004 Mar 13 & 216 \\
$1227367-031834$ & 186.90295 & -3.30937 & 291.31927 & 59.02406 & 5.422 & 1.01 & 265.19498 & -132.0/-131.8 & ECHLR & 2004 May 06 & 163 \\
$1236549-002941$ & 189.22878 & -0.49475 & 295.13962 & 62.15701 & 5.198 & 1.06 & 265.93793 & -137.9/-137.0 & ECHLR & 2004 May 06 & 190 \\
$1348366+220101$ & 207.15269 & 22.01685 & 14.56112 & 76.04436 & 5.981 & 1.02 & 270.13049 & -68.5/-75.0 & SARG & 2004 Mar 11 & 284 \\
$1407060+063311$ & 211.77515 & 6.55299 & 347.49780 & 62.68076 & 5.924 & 1.12 & 282.12589 & 2.5/-5.1 & SARG & 2004 Mar 13 & 160 \\
$1435018+070827$ & 218.75742 & 7.14080 & 358.56470 & 58.32960 & 4.856 & 1.11 & 287.82812 & 10.8/3.5 & SARG & 2004 Mar 11 & 390 \\
$1538472+494218$ & 234.69650 & 49.70488 & 79.71546 & 50.95237 & 6.076 & 1.16 & 269.88177 & -129.4/-132.1 & ECHLR & 2004 May 05 & 65 \\
\\
\multicolumn{12}{c}{\it Sgr South Leading Arm}\\
$2031334-324453$ & 307.88907 & -32.74802 & 10.20659 & -34.28811 & 11.480 & 1.04 & 20.63735 & 10.3/10.7 & MIKE & 2005 Aug 15 & 61 \\
$2037196-291738$ & 309.33173 & -29.29385 & 14.63141 & -34.68745 & 7.921 & 1.10 & 22.02419 & -82.2/-85.5 & MIKE & 2005 Aug 15 & 105 \\
$2046335-283547$ & 311.63974 & -28.59648 & 16.07145 & -36.47462 & 10.207 & 1.05 & 24.08287 & -148.4/-181.2 & MIKE & 2005 Aug 15 & 92 \\
$2050020-345336$ & 312.50839 & -34.89326 & 8.48206 & -38.45911 & 8.098 & 1.02 & 24.35087 & 9.0/7.8 & MIKE & 2005 Aug 15 & 120 \\
$2105585-275602$ & 316.49393 & -27.93392 & 18.17347 & -40.47592 & 11.639 & 1.09 & 28.40276 & 1.0/18.4 & MIKE & 2005 Aug 15 & 59 \\
$2114412-301256$ & 318.67175 & -30.21557 & 15.70152 & -42.80332 & 8.882 & 1.07 & 30.01327 & -103.6/-93.2 & MIKE & 2005 Aug 15 & 100 \\
$2130445-210034$ & 322.68533 & -21.00944 & 29.22441 & -44.07385 & 9.008 & 1.06 & 34.97513 & 179.4/156.1 & MIKE & 2005 Aug 15 & 102 \\
$2135183-203457$ & 323.82642 & -20.58247 & 30.25831 & -44.95516 & 9.048 & 1.10 & 36.11067 & -37.7/-41.7 & MIKE & 2005 Aug 15 & 81 \\
$2154471-224050$ & 328.69632 & -22.68056 & 29.25014 & -49.89746 & 8.853 & 1.04 & 40.16570 & -2.2/-3.7 & MIKE & 2005 Aug 15 & 100 \\
$2226328-340408$ & 336.63647 & -34.06901 & 11.32797 & -58.23137 & 11.555 & 1.03 & 44.12699 & -108.4/-105.8 & MIKE & 2005 Aug 15 & 94 \\
\\
\\
\\
\multicolumn{12}{c}{\it NGC Group}\\
$1033045+491604$ & 158.26884 & 49.26776 & 163.71259 & 55.44803 & 8.651 & 1.04 & 220.70378 & 148.4/135.1 & ECHLR & 2004 May 07 & 68 \\
$1041479+294917$ & 160.44971 & 29.82131 & 199.78569 & 61.47697 & 8.184 & 1.04 & 228.48373 & 44.9/49.0 & SARG & 2004 Mar 12 & 110 \\
$1051302+004400$ & 162.87578 & 0.73320 & 250.43880 & 50.95489 & 8.287 & 1.03 & 240.24094 & 28.0/28.6 & SARG & 2004 Mar 12 & 111 \\
$1115376+000800$ & 168.90674 & 0.13346 & 258.53650 & 54.52830 & 8.248 & 1.04 & 246.51361 & 67.1/67.7 & SARG & 2004 Mar 12 & 115 \\
$1214190+071358$ & 183.57918 & 7.23277 & 277.36386 & 68.24208 & 9.424 & 1.07 & 257.24124 & 283.6/293.5 & ECHLR & 2004 May 07 & 72 \\
$1257013+260046$ & 194.25543 & 26.01271 & 351.46259 & 88.32569 & 9.648 & 1.07 & 257.56802 & 106.8/104.1 & ECHLR & 2004 May 07 & 58 \\
$1343047+221636$ & 205.76953 & 22.27674 & 13.26046 & 77.31685 & 9.124 & 1.06 & 268.85614 & 153.1/144.1 & ECHLR & 2004 May 07 & 62 \\
$1412161+294303$ & 213.06714 & 29.71751 & 45.98463 & 72.06431 & 6.748 & 1.04 & 270.46027 & 92.0/90.0 & ECHLR & 2004 May 07 & 68 \\
$1424425+414932$ & 216.17723 & 41.82551 & 76.59016 & 65.93918 & 5.833 & 1.09 & 264.46732 & 75.6/73.9 & ECHLR & 2004 May 07 & 112 \\
$1429456+230043$ & 217.44019 & 23.01201 & 27.89475 & 67.39621 & 9.110 & 1.05 & 278.04666 & 236.0/229.7 & ECHLR & 2004 May 07 & 53 \\
$1513011+222640$ & 228.25456 & 22.44434 & 32.52866 & 57.63240 & 7.340 & 1.08 & 287.53125 & 226.0/226.6 & ECHLR & 2004 May 07 & 62 \\
$1536502+580017$ & 234.20917 & 58.00484 & 91.53002 & 47.79223 & 8.577 & 1.06 & 258.52972 & 78.5/79.3 & ECHLR & 2004 May 06 & 61 \\
$1545189+291310$ & 236.32875 & 29.21942 & 46.56559 & 51.84183 & 8.922 & 1.01 & 290.08469 & 113.8/108.0 & ECHLR & 2004 May 07 & 65 \\

\enddata

\end{deluxetable}

\begin{deluxetable}{cccc}
\tabletypesize{\scriptsize} \tablecaption{
Selected Iron Lines } \tablewidth{0pt} \tablehead{ \colhead{} &
\colhead{$\lambda$} &\colhead{$\chi$} &\colhead{}
 \\
 \colhead{Ion} &\colhead{(\AA\ )} &\colhead{(eV)} &
 \colhead{\it gf}
 }

\startdata
\ion{Fe}{+1} & 7443.018 & 4.186 & 1.778e-02 \\
 & 7447.384 & 4.956 & 9.752e-02 \\
 & 7461.521 & 2.559 & 2.951e-04 \\
 & 7498.530 & 4.143 & 6.457e-03 \\
 & 7507.261 & 4.415 & 1.067e-01 \\
 & 7511.015 & 4.178 & 1.538e+00 \\
 & 7531.141 & 4.371 & 4.018e-01 \\
 & 7540.430 & 2.728 & 1.514e-04 \\
 & 7547.910 & 5.100 & 7.129e-02 \\
 & 7568.894 & 4.283 & 1.507e-01 \\
 & 7583.787 & 3.018 & 1.380e-02 \\

\enddata
\end{deluxetable}

\begin{deluxetable}{cccccccccccc}
\rotate \tabletypesize{\scriptsize} \tablecaption{\ion{Fe}{+1}
Equivalent Width Measurements} \tablewidth{0pt} \tablehead{
\colhead{Star ID} &
 \colhead{7443.018\AA\ } &\colhead{7447.384\AA\ } &
 \colhead{7461.521\AA\ } &\colhead{7498.530\AA\ } &\colhead{7507.261\AA\ } &
 \colhead{7511.015\AA\ } &\colhead{7531.141\AA\ } &\colhead{7540.430\AA\ } &
 \colhead{7547.910\AA\ } &\colhead{7568.894\AA\ } &\colhead{7583.787\AA\ }}

\startdata
\multicolumn{12}{c}{\it Sgr Core} \\
$1849222-293217$ & 101.6 & 68.4 & 131.4 & 71.0 & ... & 247.7 & 170.5 & 100.1 & 38.8 & ... & ...  \\
$1853333-320146$ & 95.7 & 61.9 & 137.2 & 69.2 & 114.5 & 263.1 & 154.7 & 94.3 & 35.0 & ... & ...  \\
$1854283-295740$ & 62.6 & 39.5 & 148.5 & 43.9 & ... & 269.2 & 177.6 & 81.4 & 25.4 & ... & ...  \\
$1855341-302055$ & 90.5 & 63.1 & 132.1 & 64.7 & 121.2 & 204.8 & 159.7 & 91.7 & 46.4 & ... & ...  \\
$1855556-293316$ & 89.2 & 61.4 & 139.9 & 69.8 & ... & 242.2 & 169.0 & 97.6 & 39.6 & ... & ...  \\
$1902135-313030$ & 35.8 & 29.5 & 83.2 & 26.7 & ... & 126.6 & 85.5 & 57.5 & 17.3 & ... & ...  \\
\\
\multicolumn{12}{c}{\it Sgr North Leading Arm --- Best subsample}\\
$0919216+202305$ & 67.3 & 37.5 & 99.7 & 44.5 & 80.6 & 166.8 & 113.5 & 69.1 & 22.6 & 92.3 & 150.1  \\
$0925364+213807$ & 60.2 & ... & 99.8 & 44.5 & 77.5 & 166.6 & 113.5 & 69.0 & 30.5 & 102.4 & 141.5  \\
$1034395+245206$ & 62.7 & 43.9 & 95.7 & 34.4 & 77.1 & 167.3 & 114.3 & 69.9 & 23.0 & 96.6 & 142.2  \\
$1100516+130216$ & 46.9 & 30.7 & 83.7 & 35.1 & 66.1 & 153.8 & 99.9 & 50.7 & 16.4 & 88.9 & 129.8  \\
$1101112+191311$ & 81.9 & 59.7 & 115.8 & 54.3 & 100.2 & 184.1 & 123.1 & 90.3 & 36.9 & 116.7 & 176.2  \\
$1114573-215126$ & 48.9 & ... & 87.1 & 27.8 & 63.7 & 144.6 & 101.7 & 60.0 & 19.5 & 85.9 & 132.0  \\
$1116118-333057$ & 41.9 & 23.9 & 80.2 & 22.5 & 57.6 & 137.2 & 84.8 & 46.6 & 11.9 & 77.5 & 128.8  \\
$1140226-192500$ & 59.4 & 48.6 & 97.4 & 46.0 & 78.8 & 166.1 & 109.2 & 72.2 & 32.7 & 97.2 & 145.2  \\
$1249078+084455$ & 51.9 & 49.5 & 106.2 & 47.4 & 79.5 & 175.3 & 116.6 & 69.0 & 26.0 & 99.0 & 154.3  \\
$1318500+061112$ & 56.2 & 48.3 & 97.2 & 44.5 & 84.7 & 169.4 & 124.5 & 60.5 & 21.3 & 98.6 & 162.3  \\
$1319368-000817$ & 56.2 & 38.7 & 112.6 & 44.5 & 84.7 & 169.4 & 107.9 & 69.4 & 21.3 & 98.6 & 162.3 \\
$1330472-211847$ & 47.3 & 39.0 & 98.6 & 35.6 & 75.8 & 171.7 & 111.9 & 52.6 & 15.0 & 96.1 & 154.2 \\
$1334532+042053$ & 58.5 & 40.7 & 95.9 & 40.2 & 84.1 & 168.1 & 110.0 & 74.9 & 25.3 & 100.8 & 156.9 \\
$1411221-061013$ & 56.9 & 46.5 & 106.2 & 39.8 & 86.6 & 173.0 & 119.0 & 72.8 & 29.9 & 103.5 & 157.0 \\
$1450544+244357$ & 50.3 & 27.2 & 87.6 & 27.9 & 79.0 & 166.1 & 104.0 & 66.1 & 17.9 & 94.8 & 144.8  \\
$1456137+151112$ & 52.8 & 32.3 & 105.4 & 44.0 & 83.2 & 170.7 & 109.8 & 60.8 & 18.5 & 103.3 & 160.7  \\
$1512142-075250$ & 52.2 & 31.5 & 87.8 & 29.4 & 66.6 & 148.7 & 95.8 & 49.1 & 18.7 & 79.4 & 129.8 \\
\multicolumn{12}{c}{\it Sgr North Leading Arm --- Less Certain Subsample}\\
$1111493+063915$ & 55.0 & 32.1 & 113.5 & 40.6 & 89.0 & 168.7 & 126.0 & 73.1 & 25.2 & 100.0 & 155.8  \\
$1112480+013211$ & 75.7 & 54.1 & 104.7 & 53.2 & 99.0 & 192.4 & 126.2 & 66.0 & 29.7 & 117.7 & 155.3  \\
$1128316-031647$ & 74.0 & 51.7 & 118.8 & 57.9 & 102.1 & 195.5 & 129.2 & 81.3 & 36.2 & 113.8 & 172.3  \\
$1135388-022602$ & ... & 52.5 & 101.5 & 54.7 & 83.0 & 169.5 & 114.0 & 75.3 & 31.3 & 104.7 & 153.7  \\
$1208101-090753$ & ... & 32.2 & 109.2 & 40.7 & 84.8 & 179.1 & 114.8 & 61.2 & 18.2 & 102.7 & 159.9 \\
$1223590-073028$ & 53.9 & 41.3 & 99.1 & 38.5 & 75.7 & 164.6 & 104.9 & 62.1 & 19.8 & 95.3 & 148.2 \\
$1224255-061852$ & ... & 47.8 & 114.9 & 40.4 & 91.2 & 182.2 & 116.7 & 68.0 & 26.8 & 110.9 & 167.5 \\
$1227367-031834$ & 74.4 & 47.2 & 108.8 & 55.1 & 97.9 & 192.1 & 133.1 & 69.7 & 30.0 & 109.2 & 161.9  \\
$1236549-002941$ & 60.5 & 54.0 & 106.5 & 53.8 & 91.6 & 179.8 & 116.1 & 71.2 & 31.1 & 101.4 & 150.5 \\
$1348366+220101$ & 64.1 & 38.4 & 103.1 & 47.4 & 83.7 & 168.6 & 110.3 & 58.6 & 23.2 & 95.7 & 154.1  \\
$1407060+063311$ & 68.1 & ... & 103.8 & 33.7 & 76.8 & 174.7 & 121.3 & 63.2 & 18.2 & 100.9 & 154.2 \\
$1435018+070827$ & 49.9 & 39.6 & 98.8 & 36.5 & 70.1 & 171.0 & 107.7 & 63.8 & 19.9 & 103.8 & 153.0  \\
$1538472+494218$ & 39.1 & 24.3 & 86.6 & 27.6 & 60.1 & 148.9 & ... & 43.5 & 14.4 & 76.8 & 136.5  \\
\\
\multicolumn{12}{c}{\it Sgr South Leading Arm}\\
$2031334-324453$ & 47.1 & 26.6 & 92.5 & 29.8 & 72.4 & 214.3 & ... & 40.1 & ... & 109.7 & ...  \\
$2037196-291738$ & 83.4 & 48.6 & 131.1 & 59.1 & 101.0 & 221.3 & 157.2 & 78.2 & 27.4 & 132.7 & ...  \\
$2046335-283547$ & 52.2 & 24.6 & 104.4 & 27.3 & 76.3 & 198.2 & 119.8 & 55.4 & 10.3 & 107.3 & ...  \\
$2050020-345336$ & 58.2 & 39.1 & 99.2 & 39.8 & 93.8 & 193.8 & 122.5 & 69.9 & 15.2 & 112.9 & ...  \\
$2105585-275602$ & 54.8 & 36.0 & 116.2 & 33.9 & 96.7 & 200.8 & 117.6 & 67.0 & 21.6 & 108.4 & ...  \\
$2114412-301256$ & 46.4 & 34.8 & 91.8 & 24.7 & 80.5 & 183.2 & 114.9 & 49.6 & 16.4 & 106.5 & ...  \\
$2130445-210034$ & 37.2 & 26.3 & 112.0 & 24.0 & 78.3 & 200.8 & 123.9 & 42.0 & 11.1 & 90.3 & ...  \\
$2135183-203457$ & ... & ... & 122.0 & 59.7 & 95.7 & 204.3 & 142.5 & 70.5 & 18.8 & 119.5 & ...  \\
$2154471-224050$ & 60.2 & 38.3 & 116.7 & 42.7 & 99.8 & 207.8 & 138.6 & 67.7 & 27.1 & 119.3 & ...  \\
$2226328-340408$ & 39.3 & 19.0 & 82.3 & ... & 72.4 & 169.3 & 101.0 & 40.6 & 14.3 & 93.1 & ...  \\
\\
\multicolumn{12}{c}{\it NGC Group} \\
$1033045+491604$ & 59.9 & 38.2 & 99.2 & 44.4 & 85.8 & 178.8 & 113.3 & 56.7 & 26.7 & 97.7 & 153.1  \\
$1041479+294917$ & 45.3 & 31.1 & 84.6 & 26.7 & 61.6 & 151.7 & 96.8 & 40.0 & 10.7 & 82.1 & 133.1  \\
$1051302+004400$ & 34.8 & 20.4 & 80.9 & 24.3 & 60.1 & 143.7 & 95.2 & ... & 8.2 & 76.6 & 134.0  \\
$1115376+000800$ & 57.9 & 35.8 & 96.4 & 43.0 & 73.1 & 175.9 & 118.8 & 60.8 & 20.5 & 92.4 & 147.3  \\
$1214190+071358$ & 50.9 & 24.4 & 90.2 & 26.8 & 71.5 & 168.7 & 99.2 & 48.4 & 18.3 & 96.3 & ...  \\
$1257013+260046$ & 63.8 & 33.8 & 102.1 & 32.8 & 79.9 & 170.0 & 111.9 & 65.2 & 22.6 & 93.3 & 167.7  \\
$1343047+221636$ & 61.7 & 35.0 & 112.3 & 28.6 & 84.0 & 201.1 & 119.1 & 55.4 & 19.4 & 107.4 & 190.9  \\
$1412161+294303$ & 75.4 & 47.0 & 121.0 & 54.5 & 103.8 & 191.0 & 142.0 & 76.7 & 33.9 & 116.3 & 171.7  \\
$1424425+414932$ & 68.2 & 39.4 & 105.9 & 44.9 & 84.6 & 171.5 & 118.7 & 67.5 & 21.8 & 100.6 & 154.9 \\
$1429456+230043$ & 71.3 & 32.0 & 102.3 & 35.8 & 83.4 & 182.6 & 125.4 & 55.9 & 24.9 & 104.1 & 174.1 \\
$1513011+222640$ & 44.0 & 36.3 & 85.3 & 39.0 & 72.2 & 146.3 & 99.7 & 52.5 & 20.1 & 88.8 & 127.4  \\
$1536502+580017$ & 65.0 & 44.2 & 95.9 & 42.4 & 87.2 & 164.0 & 116.1 & 58.9 & 20.3 & 103.3 & 156.0 \\
$1545189+291310$ & 50.9 & 40.9 & 90.2 & 34.9 & 72.5 & 168.0 & 106.0 & 53.9 & 18.9 & 97.0 & 144.6 \\
\\
\multicolumn{12}{c}{\it Calibration Stars} \\
Arcturus (SARG) & 63.2 & 52.6 & 88.2 & 38.9 & 89.6 & 192.1 & 122.2 & 62.4 & 23.4 & 110.7 & 154.2 \\
Arcturus (KPNO) & 69.3 & 46.9 & 94.3 & 52.4 & 82.8 & 154.1 & 103.8 & 44.7 & 24.8 & 105.1 & 152.4 \\
Arcturus (MIKE) & 60.1 & 41.5 & 90.9 & 49.0 & 92.0 & 182.6 & 120.1 & 55.9 & 27.4 & 110.5 & 151.9 \\
$\alpha$ Tau & ... & 71.0 & ... & ... & 113.0 & ... & ... & ... & ... & ... & 175.0 \\
$\beta$ And & 66.1 & 51.7 & 119.3 & 57.6 & 102.8 & 205.2 & 138.7 & 76.2 & 36.0 & 125.3 & 184.7 \\
$\rho$ Per & 64.6 & 51.1 & 107.3 & 50.6 & 86.4 & 180.9 & 103.0 & 64.5 & 35.2 & 112.8 & 148.0 \\
$\beta$ Peg & 63.4 & 49.2 & 113.2 & 50.9 & 83.3 & 180.8 & 131.9 & 71.7 & 26.4 & 105.5 & 168.5 \\

\enddata

\end{deluxetable}

\begin{deluxetable}{cccccccc}
\tabletypesize{\scriptsize} \tablecaption{
Bright Red Giant Standard Stars } \tablewidth{0pt} \tablehead{ \colhead{Star} &
\colhead{$\pi$(mas)} &\colhead{$d$(pc)} &\colhead{$M_{\rm V}$} &
 \colhead{$M_{\rm K}$} &\colhead{$M_{\rm bol}$} &\colhead{$T_{\rm eff}$($V-K$)} &
 \colhead{$T_{\rm eff}$($J-K$)}
 }

\startdata
$\alpha$ Tau & 50.1$\pm$1.0 & 20$\pm$0.4   & -0.74$\pm$0.05 & -4.41$\pm$0.05 & -1.82$\pm$0.07 & 3900 & 3950 \\
$\beta$ And   & 16.4$\pm$0.8 & 61$\pm$3.0   & -1.88$\pm$0.10 & -5.74$\pm$0.10 & -3.06$\pm$0.11 & 3800 & 3850 \\
$\rho$ Per      & 10.0$\pm$0.8 & 100$\pm$8.0 & -1.58$\pm$0.17 & -6.90$\pm$0.17 & -4.00$\pm$0.18 & 3500 & 3650 \\
$\beta$ Peg   & 16.4$\pm$0.7 &  61$\pm$2.6  & -1.52$\pm$0.09 & -6.15$\pm$0.09 & -3.35$\pm$0.10 & 3600 & 3750 \\

\enddata
\end{deluxetable}

\begin{deluxetable}{cccccccccc}
\rotate
\tabletypesize{\scriptsize} \tablecaption{ Derived
Parameters for Red Giant Standard Stars } \tablewidth{0pt}
\tablehead{ \colhead{Star} & \colhead{$T_{\rm eff}$}
&\colhead{$\log$($L$/L$_{\odot}$)}
&\colhead{M$_{\odot}$($Z$=0.019)} &
 \colhead{$\log g$ ($Z$=0.019)} &\colhead{M$_{\odot}$($Z$=0.008)} &\colhead{$\log g$ ($Z$=0.008)}&
 \colhead{$\log g$(final)} &\colhead{[Fe/H](final)} &\colhead{$\xi$(km-s$^{-1}$)}
 }

\startdata
$\alpha$ Tau &  3925$\pm$75 & 2.61$\pm$0.03 & 1.5$\pm$0.3 & 1.33$\pm$0.08 & 0.9$\pm$0.2 & 1.11$\pm$0.09 & 1.35 & +0.06$\pm$0.12 & 1.5 \\
$\beta$ And   &  3825$\pm$75 & 3.10$\pm$0.04 & 2.0$\pm$0.3 & 0.92$\pm$0.07 & 1.5$\pm$0.3 & 0.80$\pm$0.09 & 0.88 & -0.15$\pm$0.06 & 1.7 \\
$\rho$ Per      &  3575$\pm$75 & 3.48$\pm$0.07 & 2.0$\pm$0.3 & 0.42$\pm$0.07 & 1.5$\pm$0.3 & 0.30$\pm$0.11 & 0.40 & -0.08$\pm$0.12 & 1.5  \\
$\beta$ Peg   &  3675$\pm$75 & 3.22$\pm$0.04 & 2.0$\pm$0.3 & 0.73$\pm$0.09 & 1.3$\pm$0.3 & 0.54$\pm$0.10 & 0.60 & -0.33$\pm$0.08 & 1.6  \\

\enddata
\end{deluxetable}

\begin{deluxetable}{rcccc}
\tabletypesize{\scriptsize} \tablecaption{ Red Giant Standard Star
Parameters from Isochrones } \tablewidth{0pt} \tablehead{
\colhead{Star} & \colhead{$T_{\rm eff}$} &\colhead{$\log g$}
&\colhead{[Fe/H]} &\colhead{$\xi$(km-s$^{-1}$)}
 }

\startdata
$\alpha$ Tau (2.5 Gyr Y$^{2}$) & 3950 & 1.4 & -0.04 & 1.9 \\
 (2.5 Gyr Girardi)                           & 3950 & 1.3 & -0.06 & 1.9 \\
 (1.0 Gyr Girardi)                           & 3950 & 1.3 & -0.06 & 1.9 \\
 $\beta$ And (2.5 Gyr Y$^{2}$)  & 3850 & 0.9 & -0.33 & 2.0 \\
  (2.5 Gyr Girardi)                          & 3850 & 0.9 & -0.33 & 2.0 \\
  (1.0 Gyr Girardi)                          & 3850 & 0.9 & -0.33 & 2.0 \\
  $\rho$ Per (2.5 Gyr Y$^{2}$)    & 3650 & 0.8 & -0.04 & 1.3 \\
  (2.5 Gyr Girardi)                          & 3650 & 0.7 & -0.09 & 1.4 \\
  (1.0 Gyr Girardi)                          & 3650 & 0.8 & -0.04 & 1.3 \\
  $\beta$ Peg (2.5 Gry Y$^{2}$) & 3750 & 0.6 & -0.47 & 1.7 \\
  (2.5 Gyr Girardi)                          & 3750 & 0.6 & -0.47 & 1.7 \\
  (1.0 Gyr Girardi)                          & 3750 & 0.5 & -0.51 & 1.7 \\

\enddata
\end{deluxetable}

\begin{deluxetable}{cccccc}
\tabletypesize{\scriptsize} \tablecaption{Derived Stellar
Parameters for The Program Stars} \tablewidth{0pt} \tablehead{
\colhead{Star ID.} & \colhead{$T_{\rm eff}$} & \colhead{$\log g$} & \colhead{$\xi$ }  & \colhead{[Fe/H]}
& \colhead{standard}\\
\colhead{} & \colhead{(K)} & \colhead{(${\rm cm\,
s^{-2}}$)} & \colhead{(${\rm km\,s^{-1}}$)} & \colhead{} &
\colhead{deviation}}

\startdata
\multicolumn{6}{c}{\it Sgr Core}  \\
$1849222-293217$ & 3850 & 0.9 & 2.43 & -0.20 & 0.10 \\
$1853333-320146$ & 3750 & 0.7 & 2.60 & -0.30 & 0.14 \\
$1854283-295740$ & 3750 & 0.0(-) & 3.21 & -0.97 & 0.06 \\
$1855341-302055$ & 3800 & 1.0 & 1.84 & 0.02 & 0.08 \\
$1855556-293316$ & 3700 & 0.5 & 2.36 & -0.27 & 0.07 \\
$1902135-313030$ & 3750 & 0.0(-) & 1.04 & -1.04 & 0.11 \\
\\
\multicolumn{6}{c}{\it Sgr North Leading Arm --- Best Subsample}\\
$0919216+202305$ & 3700 & 0.25 & 1.47 & -0.63 & 0.08 \\
$0925364+213807$ & 3600 & 0.5 & 1.29 & -0.23 & 0.07 \\
$1034395+245206$ & 3700 & 0.25 & 1.45 & -0.65 & 0.09 \\
$1100516+130216$ & 3800 & 0.0 & 1.35 & -1.06 & 0.08 \\
$1101112+191311$ & 3700 & 0.8 & 1.51 & 0.02 & 0.09 \\
$1114573-215126$ & 3550 & 0.0(-) & 1.33 & -0.81 & 0.07 \\
$1116118-333057$ & 3650 & 0.0(-) & 1.39 & -1.13 & 0.05 \\
$1140226-192500$ & 3800 & 0.6 & 1.16 & -0.38 & 0.05 \\
$1249078+084455$ & 3800 & 0.3 & 1.52 & -0.67 & 0.10 \\
$1318500+061112$ & 3850 & 0.4 & 1.67 & -0.78 & 0.10 \\
$1319368-000817$ & 3500 & 0.0(-) & 1.64 & -0.59 & 0.08 \\
$1330472-211847$ & 3850 & 0.0(-) & 1.75 & -1.10 & 0.08 \\
$1334532+042053$ & 3700 & 0.25 & 1.49 & -0.62 & 0.06 \\
$1411221-061013$ & 3700 & 0.25 & 1.51 & -0.56 & 0.06 \\
$1450544+244357$ & 3800 & 0.0 & 1.63 & -1.08 & 0.08 \\
$1456137+151112$ & 3750 & 0.0(-) & 1.71 & -0.98 & 0.08 \\
$1512142-075250$ & 3700 & 0.0(-) & 1.26 & -0.97 & 0.08 \\
\multicolumn{6}{c}{\it Sgr North Leading Arm --- Less Certain Subsample}\\
$1111493+063915$ & 3600 & 0.0(-) & 1.71 & -0.70 & 0.09 \\
$1112480+013211$ & 3800 & 0.5 & 1.60 & -0.49 & 0.13 \\
$1128316-031647$ & 3700 & 0.9 & 1.64 & -0.04 & 0.05 \\
$1135388-022602$ & 3700 & 0.9 & 1.23 & 0.00 & 0.07 \\
$1208101-090753$ & 3750 & 0.0(-) & 1.82 & -0.99 & 0.07 \\
$1223590-073028$ & 3600 & 0.0 & 1.50 & -0.72 & 0.08 \\
$1224255-061852$ & 3750 & 0.3 & 1.71 & -0.65 & 0.08 \\
$1227367-031834$ & 3850 & 0.5 & 1.68 & -0.55 & 0.08 \\
$1236549-002941$ & 3750 & 0.5 & 1.33 & -0.39 & 0.10 \\
$1348366+220101$ & 3800 & 0.1 & 1.49 & -0.82 & 0.09 \\
$1407060+063311$ & 3700 & 0.0(-) & 1.78 & -0.95 & 0.09 \\
$1435018+070827$ & 3700 & 0.0(-) & 1.61 & -0.92 & 0.09 \\
$1538472+494218$ & 3600 & 0.0(-) & 1.52 & -1.06 & 0.08 \\
\\
\multicolumn{6}{c}{\it Sgr South Leading Arm}   \\
$2031334-324453$ & 3800 & 0.0(-) & 2.67 & -1.32 & 0.11 \\
$2037196-291738$ & 3700 & 0.0 & 2.32 & -0.70 & 0.09 \\
$2046335-283547$ & 3750 & 0.0(-) & 2.56 & -1.26 & 0.06 \\
$2050020-345336$ & 3800 & 0.0(-) & 2.12 & -1.04 & 0.10 \\
$2105585-275602$ & 3700 & 0.0(-) & 2.13 & -0.96 & 0.10 \\
$2114412-301256$ & 3750 & 0.0(-) & 2.06 & -1.15 & 0.10 \\
$2130445-210034$ & 3750 & 0.0(-) & 2.62 & -1.35 & 0.10 \\
$2135183-203457$ & 3700 & 0.0(-) & 2.30 & -0.90 & 0.13 \\
$2154471-224050$ & 3800 & 0.1 & 2.14 & -0.91 & 0.06 \\
$2226328-340408$ & 3800 & 0.0(-) & 1.93 & -1.34 & 0.09 \\
\\
\multicolumn{6}{c}{\it NGC Group}    \\
$1033045+491604$ & 3800 & 0.3 & 1.56 & -0.75 & 0.09 \\
$1041479+294917$ & 3800 & 0.0(-) & 1.56 & -1.24 & 0.10 \\
$1051302+004400$ & 3800 & 0.0(-) & 1.65 & -1.38 & 0.06 \\
$1115376+000800$ & 3800 & 0.0 & 1.56 & -0.96 & 0.10 \\
$1214190+071358$ & 3750 & 0.0(-) & 1.73 & -1.13 & 0.10 \\
$1257013+260046$ & 3750 & 0.0(-) & 1.68 & -0.96 & 0.09 \\
$1343047+221636$ & 3750 & 0.0(-) & 2.26 & -1.08 & 0.10 \\
$1412161+294303$ & 3800 & 0.6 & 1.76 & -0.43 & 0.07 \\
$1424425+414932$ & 3700 & 0.1 & 1.59 & -0.72 & 0.07 \\
$1429456+230043$ & 3750 & 0.0(-) & 1.88 & -0.97 & 0.11 \\
$1513011+222640$ & 3700 & 0.0(-) & 1.17 & -0.85 & 0.10 \\
$1536502+580017$ & 3750 & 0.0(-) & 1.53 & -0.84 & 0.10 \\
$1545189+291310$ & 3850 & 0.1 & 1.52 & -1.00 & 0.09 \\

\enddata

\end{deluxetable}

\begin{figure}
\plotone{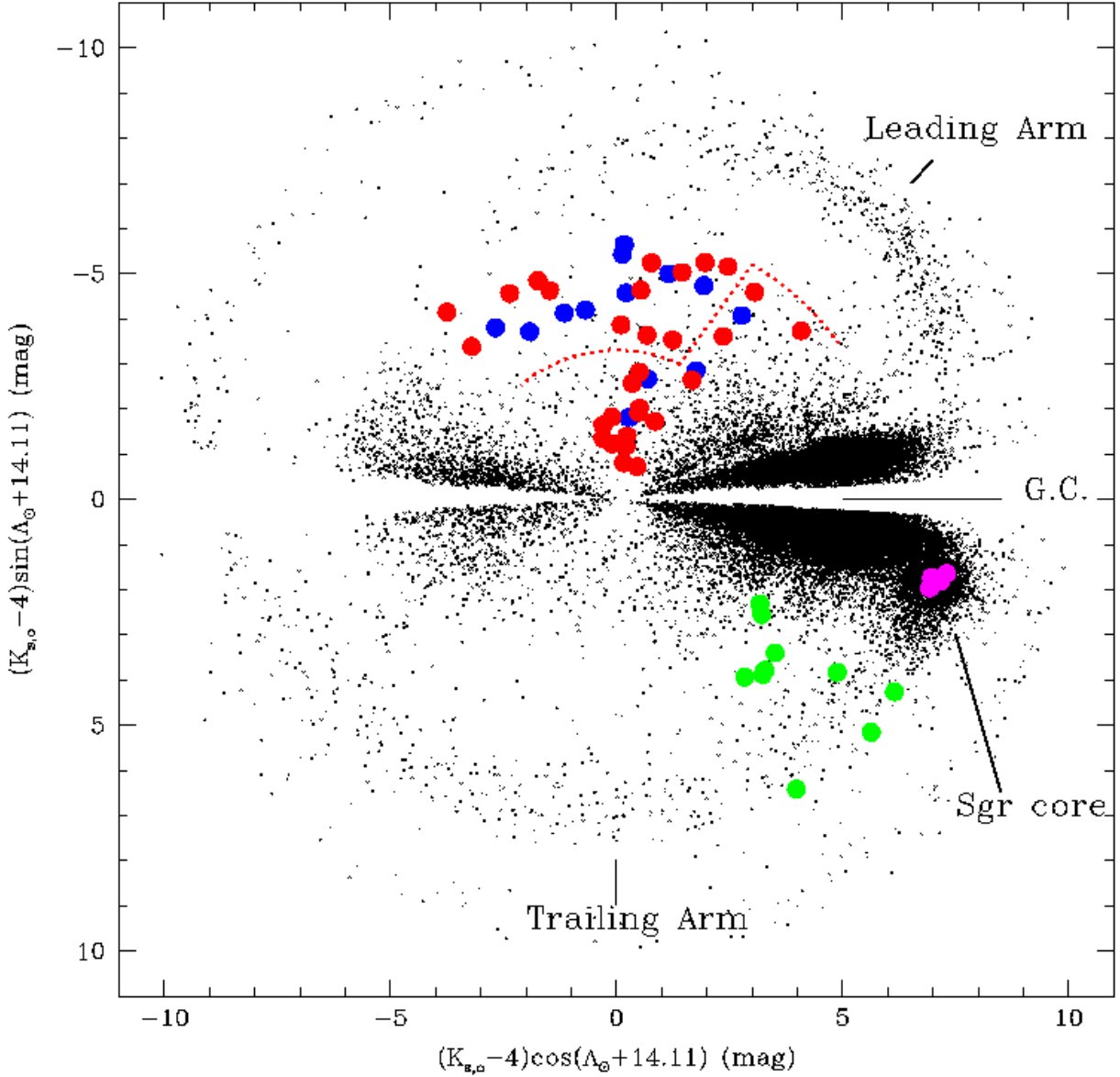} \caption{Sgr orbital plane position of M giants
lying within 10$^{\circ}$ of that nearly polar plane and having
$(J-K_s)_o > 1.0$ and $E(B-V)<0.555$ (black dots). The Sun lies at
the origin of the distribution and stars are positioned in a polar
projection based on their dereddened 2MASS $K_s$ magnitude (radial
direction, after subtraction of 4 mag) and angle from the center
of Sgr, $\Lambda_{\sun}$, increasing in the direction of the
trailing arm (i.e. counterclockwise).  The term
($\Lambda_{\sun}+14.11$) places the intersection of the Sgr and
Galactic planes horizontal across the center of the figure (see
Paper I for a further description of this coordinate system).
Stars from the four subsamples for which we present  new data here
are represented by the large colored symbols: Sgr core (magenta),
leading arm north (red), leading arm south (green) and the ``NGC"
group of stars having positive GSR radial velocities off the main
leading arm trend (blue).  We do not show the positions of other
stars in the Sgr core for which data are taken from the
literature, but these stars lie near the magenta points in the
figure.  The red dotted line delineates the division of the north
leading arm into the ``best" (beyond the dotted line) and ``less
certain" (inside the dotted line) subsamples.
 }
\end{figure}

\begin{figure}
\plotone{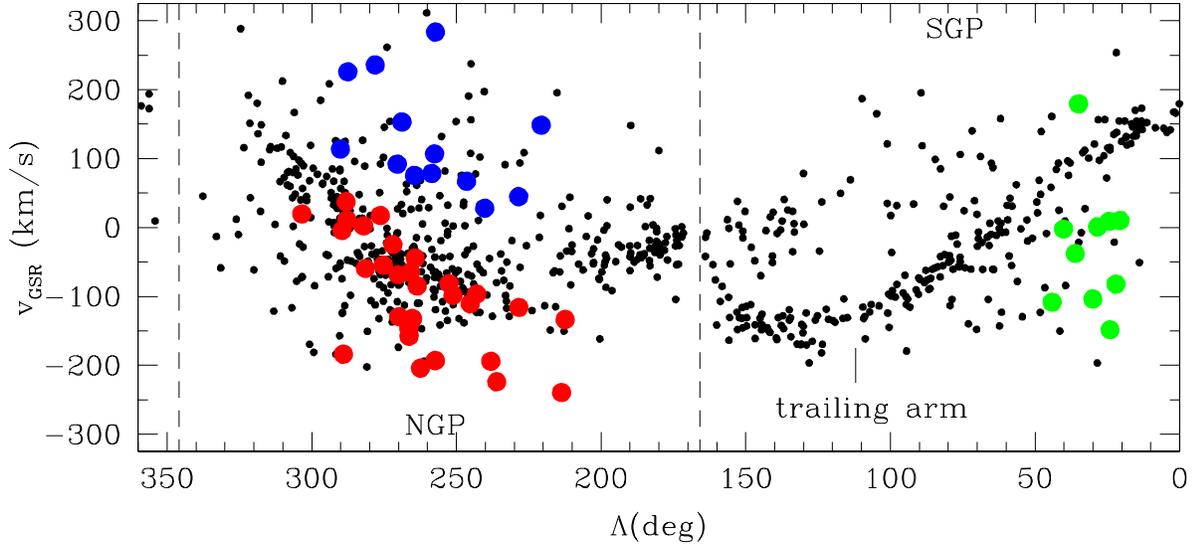} \caption{GSR radial velocities of stars as a
function of their $\Lambda_{\sun}$ angle from Sgr center from data
obtained in our ongoing medium resolution spectroscopic study. For
clarity, black dots show only $(J-K_s)_o > 1.0$ M giants with
projected distances less than 5 kpc from the Sgr orbital mid-plane
and closer than 50 kpc to the Sun.  The approximate
$\Lambda_{\sun}$ positions of the North Galactic Pole (NGP) and
South Galactic Pole (SGP) are indicated (though these actual
points on the celestial sphere actually lie approximately
$13^{\circ}$ off the Sgr plane being shown), as are the positions
of the Galactic plane (dashed lines). The coherent velocity
sequence of the Sgr trailing arm, not studied here, is also
indicated. Stars from the four subsamples for which we present new
data here are represented by the same colored symbols as in Fig.\
1.
 }
\end{figure}

\begin{figure}
\plotone{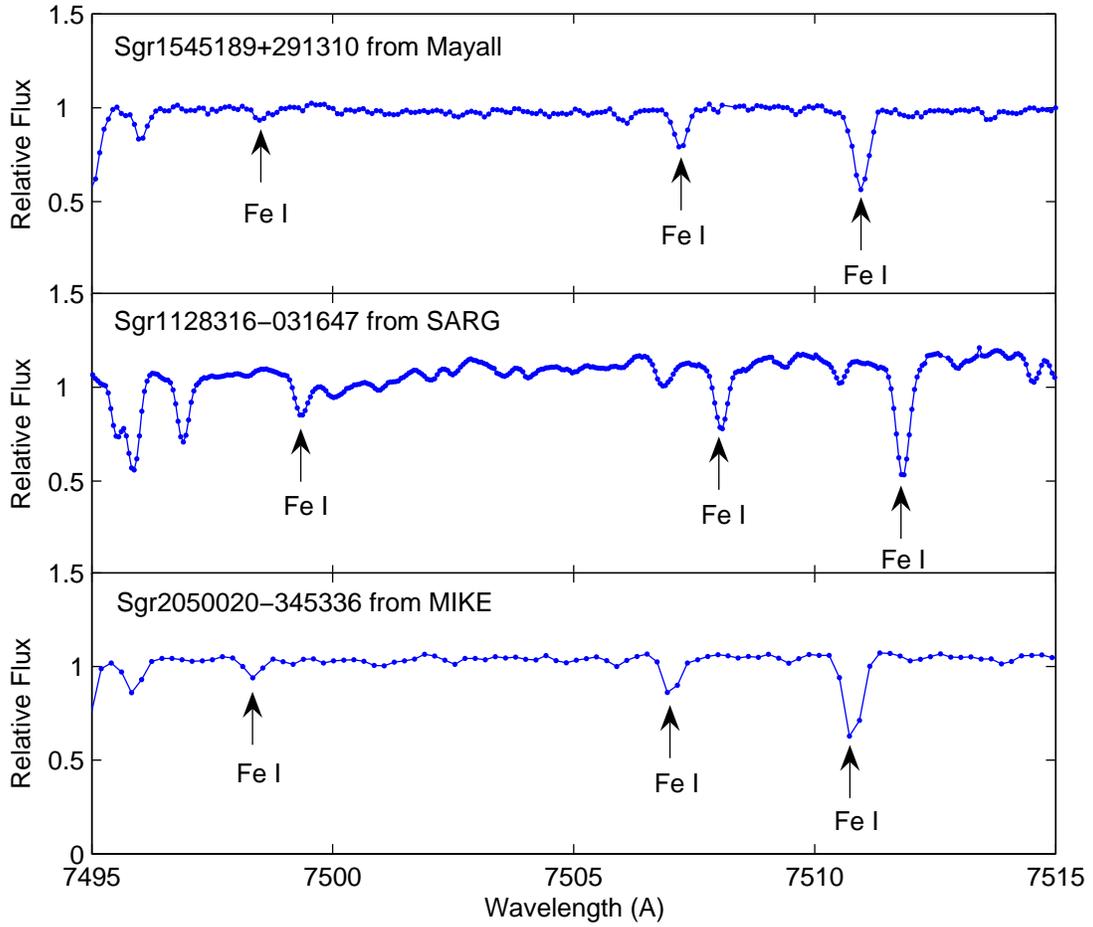} \caption{ Sample spectra of three M giants from
the three different spectrographs used in this study. Three sample
iron lines are identified in the figure.
 }
\end{figure}

\begin{figure}
\plotone{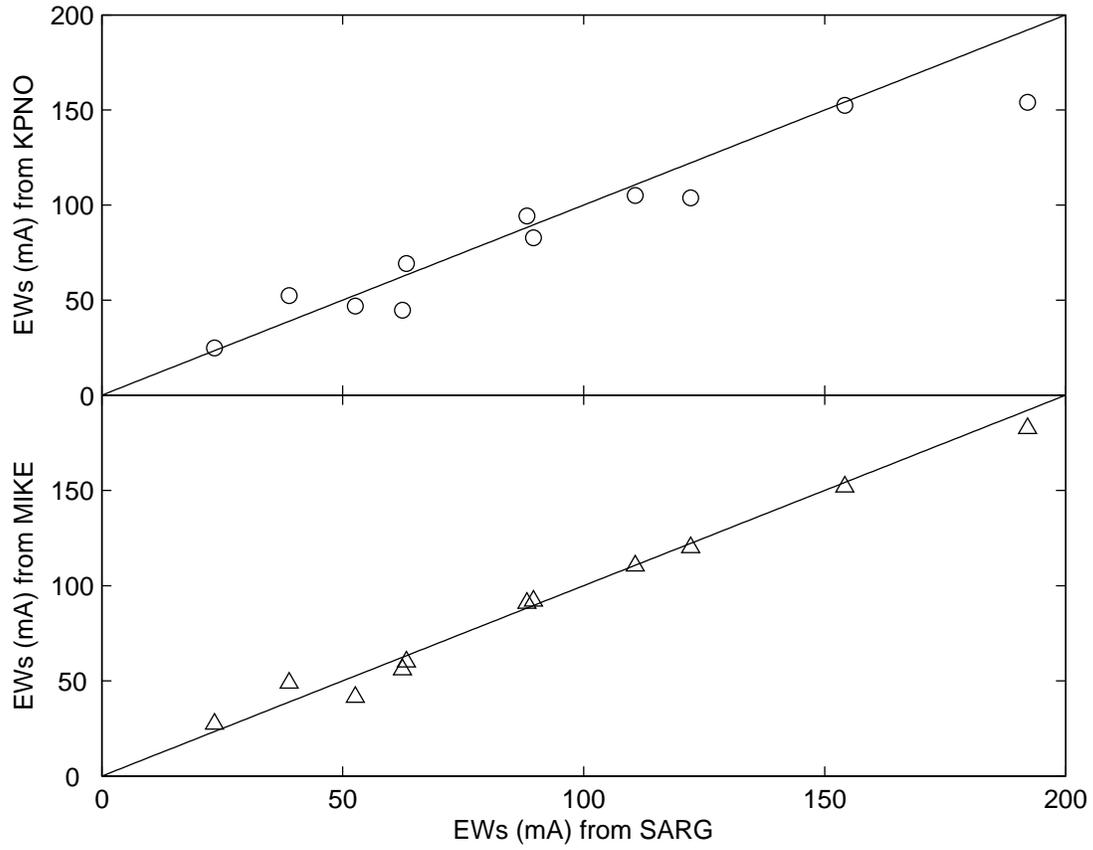} \caption{Comparison of \ion{Fe}{+1} line EWs
among the Mayall, SARG and MIKE echelles as measured in the star
Arcturus. The mean offset between Mayall and SARG is $11.00 \pm
10.65$ m\AA\, and $4.91 \pm 3.76$ m\AA\ between MIKE and SARG.
 }
\end{figure}

\begin{figure}
\plotone{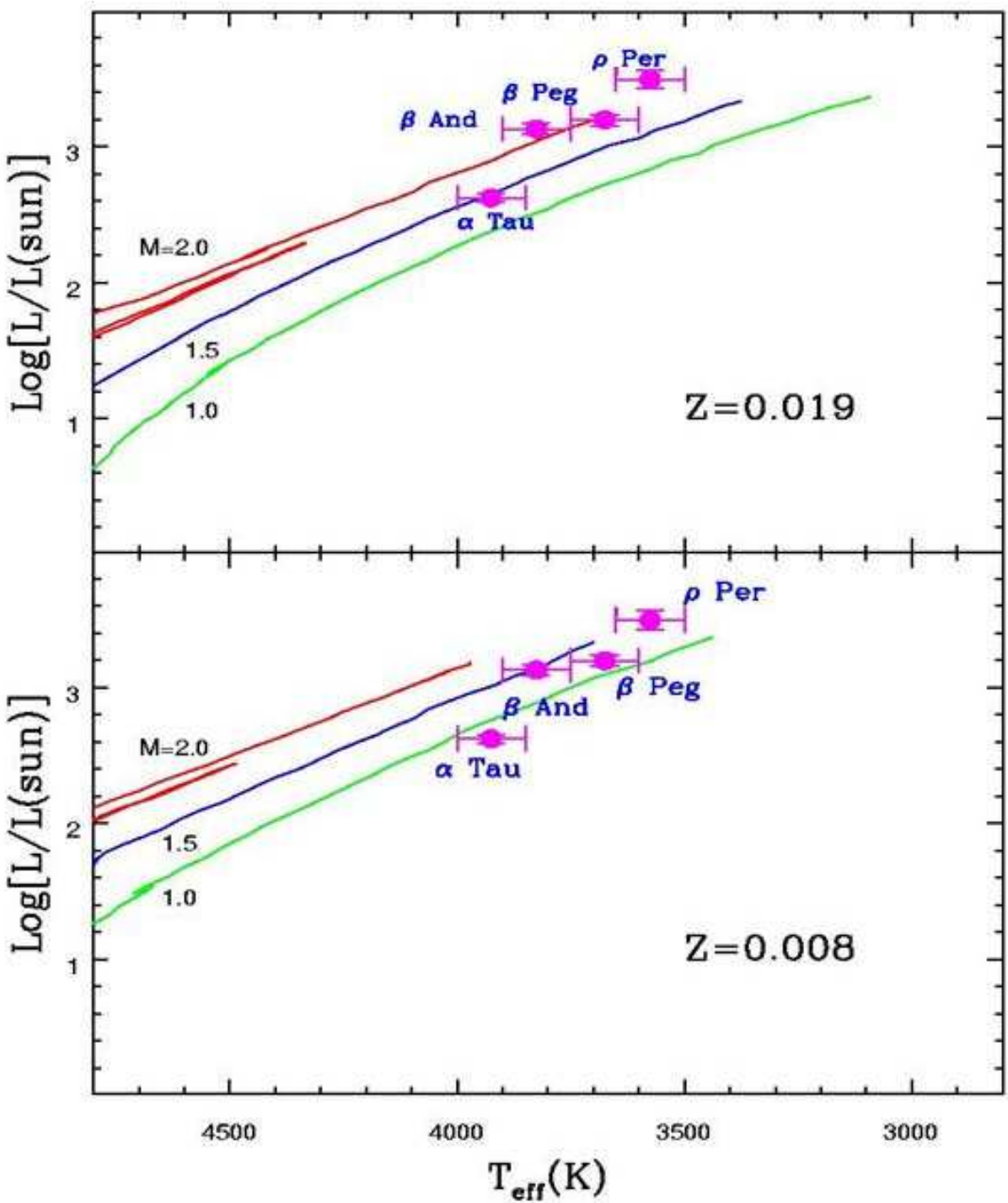} \caption{ Stellar luminosities versus effective
temperatures for the four standard red giants are calculated by
assuming $M_{\rm bol}$=4.74 for the Sun. Also plotted in this
figure are stellar model tracks from the Padua grid for masses of
$M$=1.0, 1.5, and 2.0M$_{\odot}$. The top panel shows models with
near-solar metallicity ($Z$=0.019), while the bottom panel has
models with [M/H]$\sim$-0.4 ($Z$=0.008).
 }
\end{figure}

\begin{figure}
\plotone{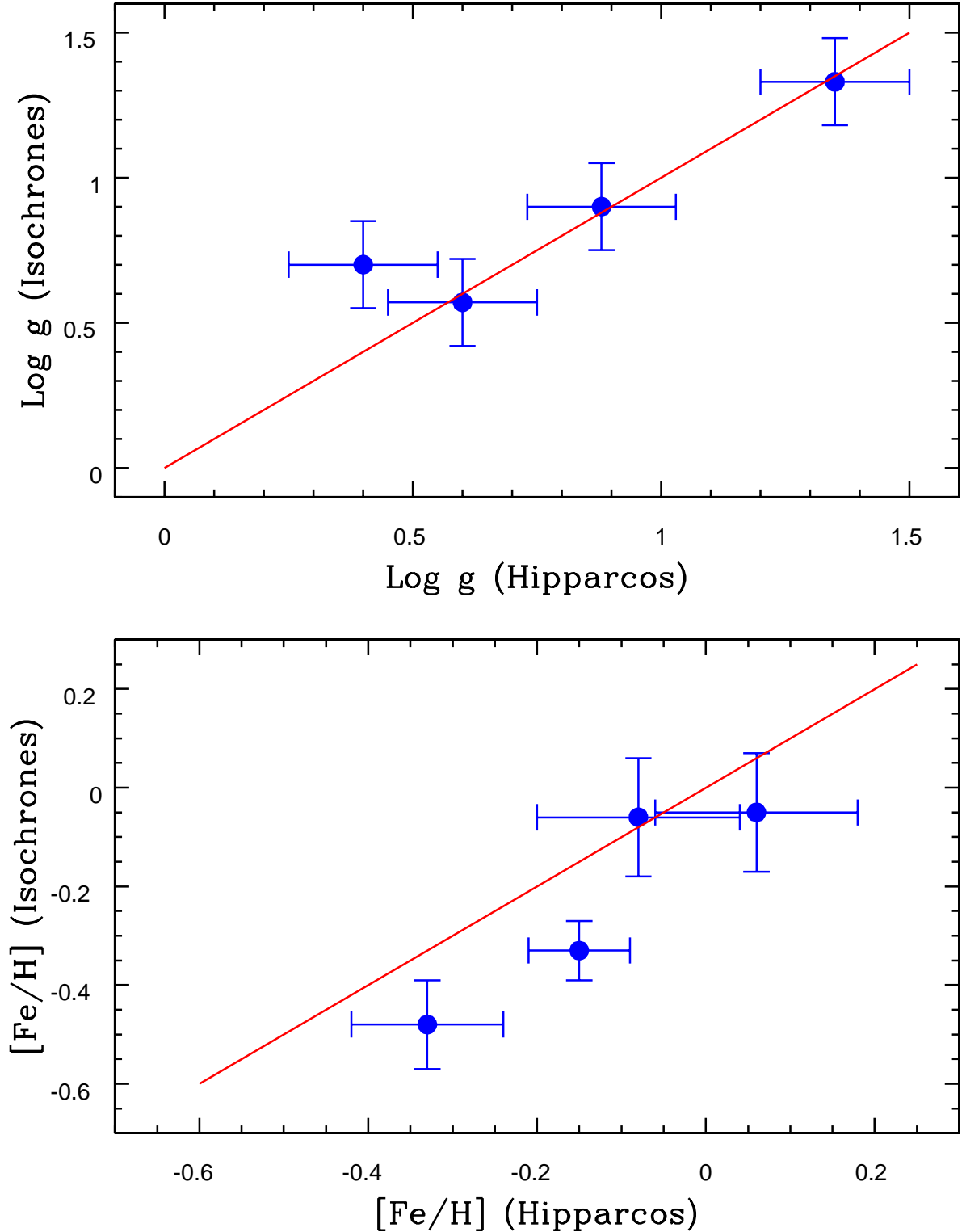} \caption{ A graphical comparison of surface
gravities (top panel) and [Fe/H] values (bottom panel) derived
from the Hipparcos parallaxes and the isochrone method for the
four bright standard stars analyzed in \S4.3.
 }
\end{figure}

\begin{figure}
\plotone{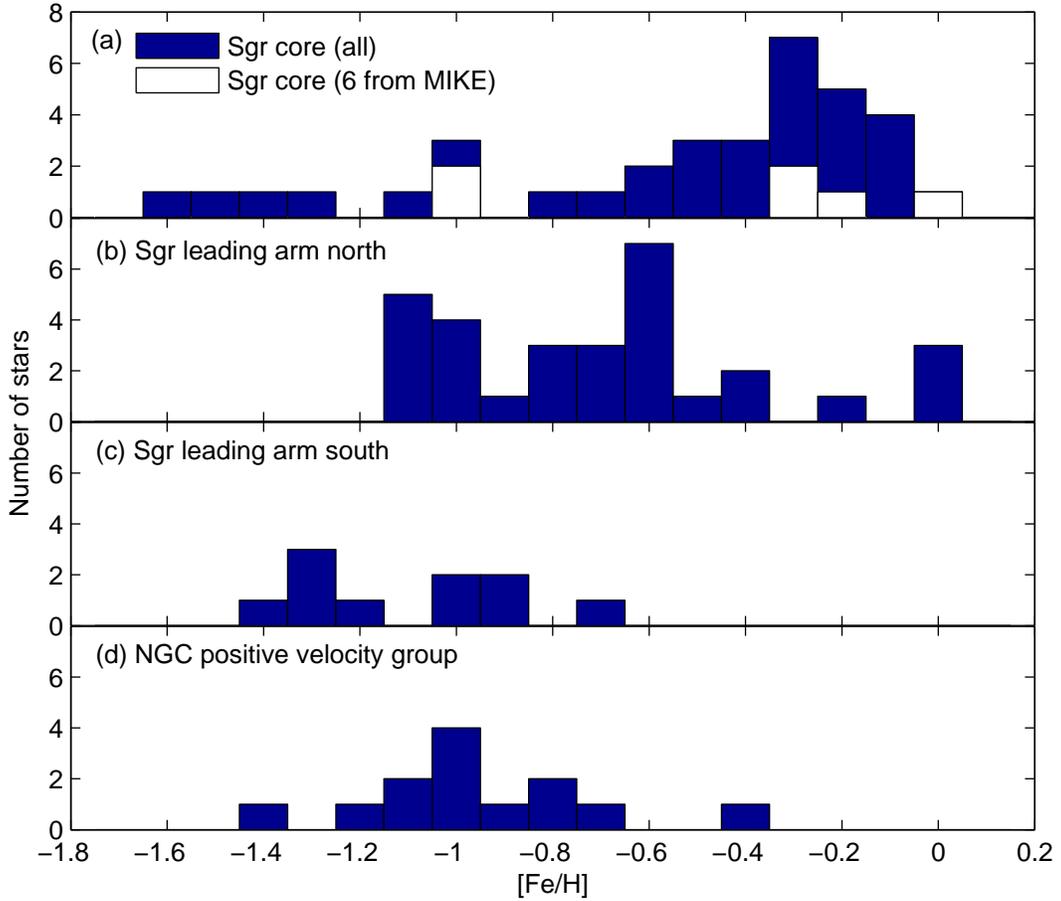} \caption{The MDF derived for stars in the (a) Sgr
core (including all previous echelle data (Smecker-Hane \&
McWilliam 2002; Monaco et al. 2005) with our six newly observed
stars shown in white), (b) leading arm north of the Sun, (c)
leading arm south of the Sun, and (d) the positive-velocity, NGC
moving group (blue circles in Fig.\ 1). }
\end{figure}

\begin{figure}
\plotone{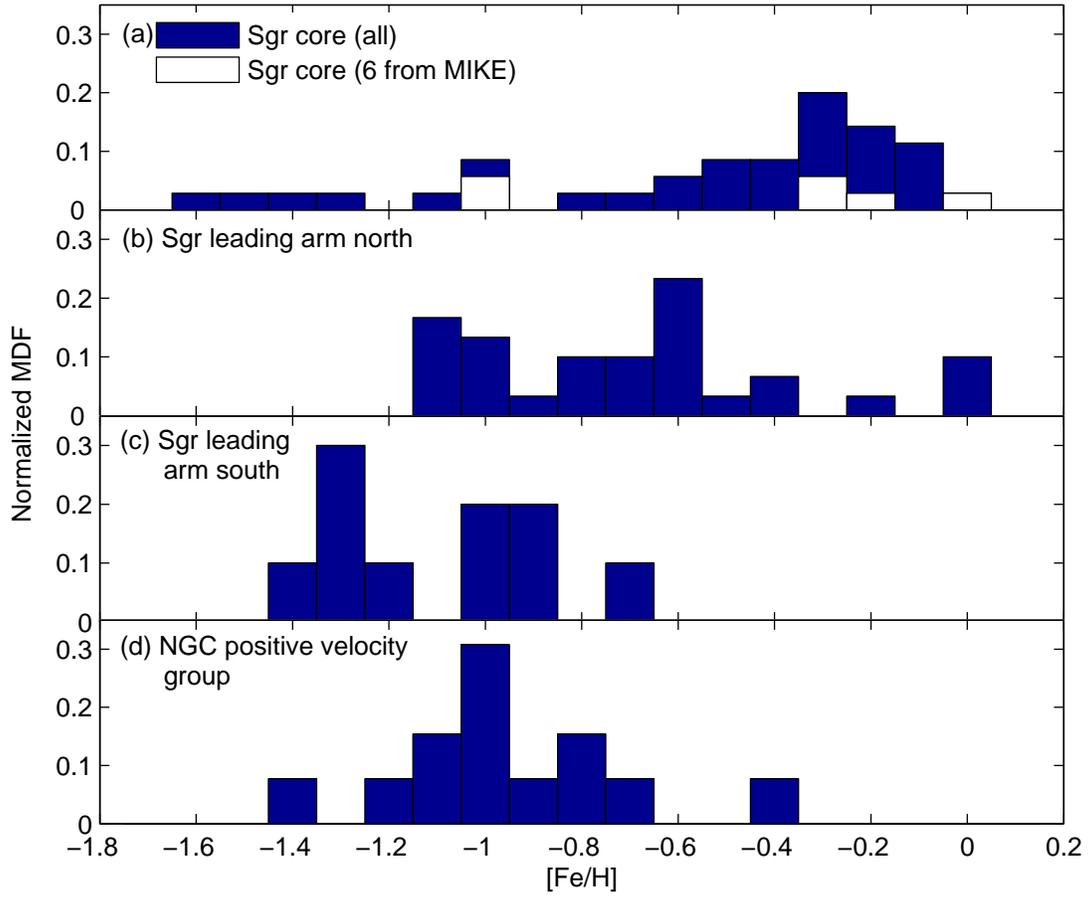} \caption{Same as Figure 7, but with each panel
scaled to the same ``normalized MDF" scale.}
\end{figure}

\begin{figure}
\plotone{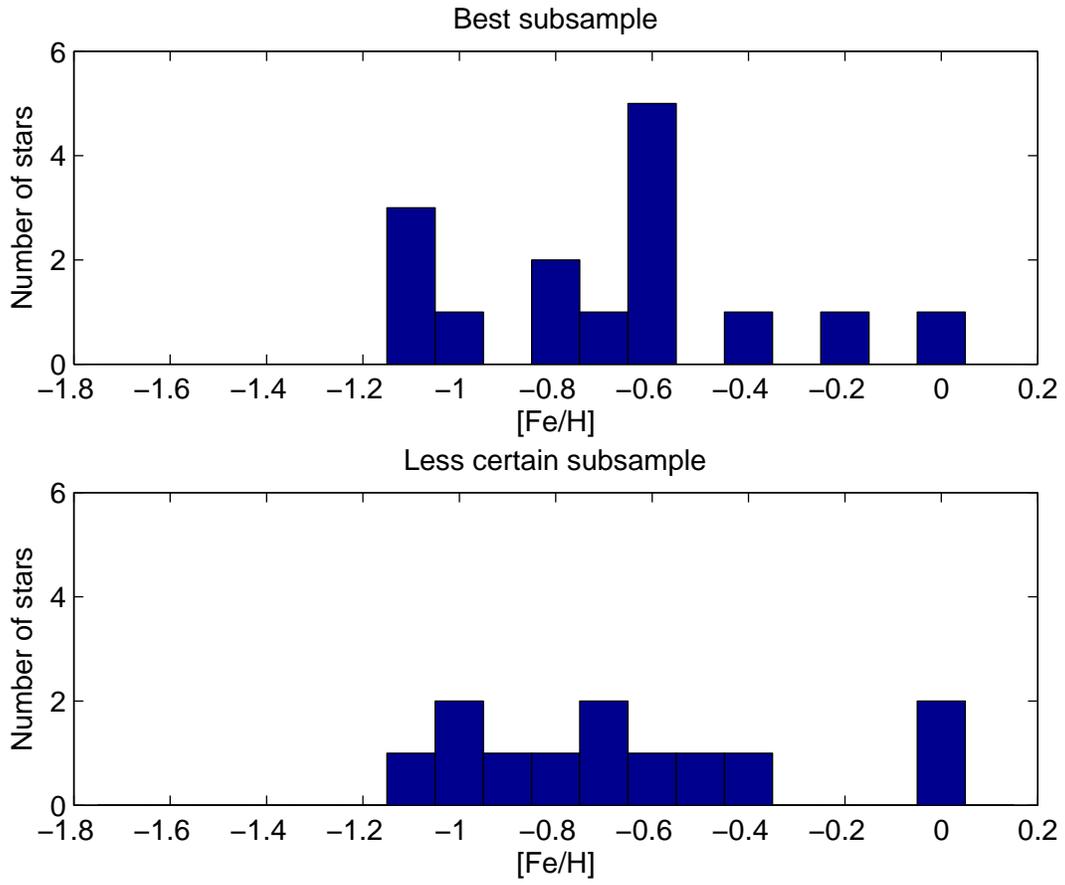} \caption{MDF comparison of subsamples of
``leading arm north" stars, divided into the ``best" (generally
farther) and ``less certain" (generally closer) Sgr stream groups.
}
\end{figure}

\begin{figure}
\plotone{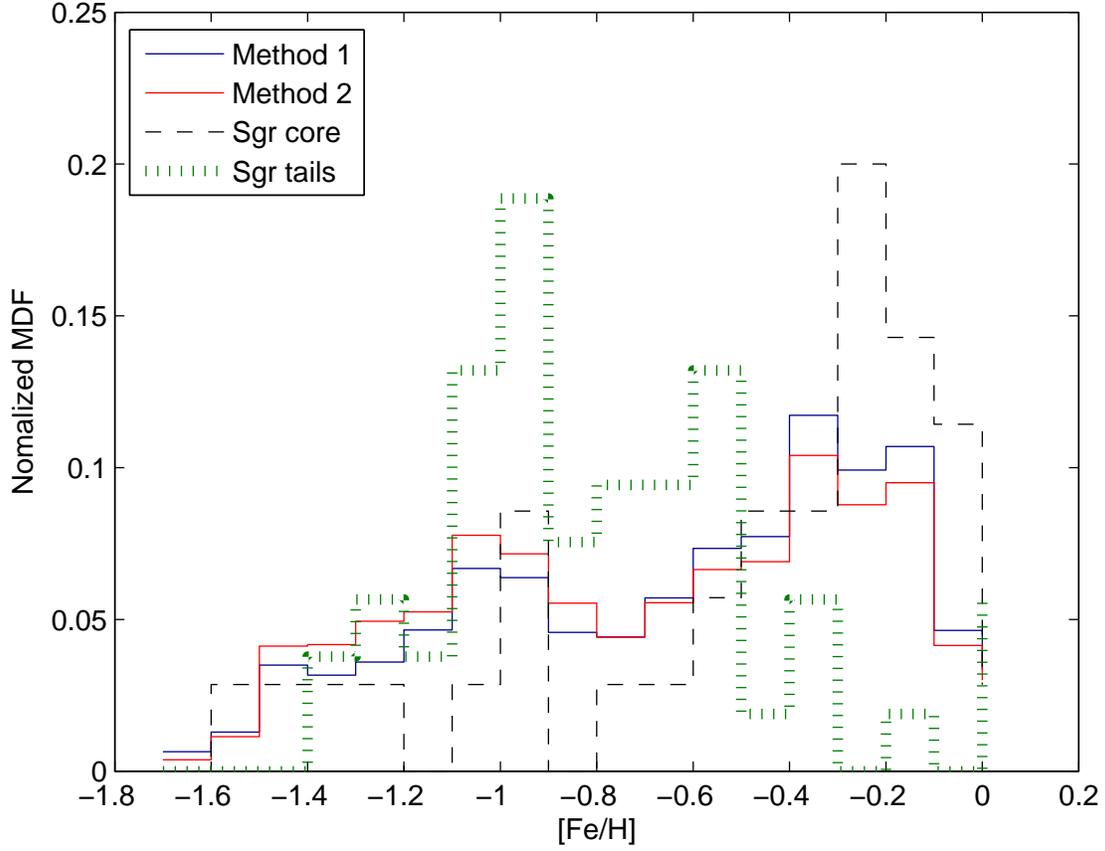} \caption{The approximate MDF of Sgr several Gyr
ago estimated from linear combinations of those shown in Fig.\ 8
by the two methods described in \S7 (blue lines for the first
method, and red lines for the second method). The histograms have
been boxcar-smoothed with a 3 bin kernal. The MDF of the Sgr core
(dashed lines) and all tail stars (Figs.\ 8b-d, green dotted
lines) are shown for comparison.}
\end{figure}

\end{document}